\documentclass[aps,pre,twocolumn,groupedaddress,showpacs]{revtex4}

\pdfoutput=1

\usepackage[latin1]{inputenc}
\usepackage[english]{babel}
\usepackage{setspace}
\usepackage{graphicx}
\usepackage{amssymb}
\usepackage{amsmath}
\usepackage{amscd}
\usepackage{amstext}
\usepackage{rotating}
\usepackage{color}
\usepackage{epstopdf}
\usepackage{times}

\begin{document}

\title{Optimized broad-histogram simulations for strong first-order phase transitions:\\
         Droplet transitions in the large-Q Potts model}

\author{Bela Bauer$^1$, Emanuel Gull$^2$, Simon Trebst$^3$, Matthias Troyer$^1$, David A. Huse$^4$}
\address{$^1$Theoretische Physik, ETH Zurich, 8093 Zurich, Switzerland}
\address{$^2$ Department of Physics, Columbia University, New York, NY 10027, USA}
\address{$^3$Microsoft Research, Station Q, University of California, Santa Barbara, CA 93106, USA}
\address{$^4$Department of Physics, Princeton University, Princeton, NJ 08544, USA}

\date{\today}

\begin{abstract}
The numerical simulation of strongly first-order phase transitions has remained a notoriously
difficult problem even for classical systems due to the exponentially suppressed (thermal)
equilibration in the vicinity of such a transition.
In the absence of efficient update techniques, a common approach to improve equilibration
in Monte Carlo simulations is to broaden the sampled statistical ensemble beyond the bimodal
distribution of the canonical ensemble.
Here we show how a recently developed feedback algorithm can systematically optimize
such broad-histogram ensembles and significantly speed up equilibration in comparison
with other extended ensemble techniques such as flat-histogram, multicanonical or
Wang-Landau sampling.
As a prototypical example of a strong first-order transition we simulate the two-dimensional
Potts model with up to $Q=250$ different states on large systems.
The optimized histogram develops a distinct multipeak structure, thereby resolving entropic barriers
and their associated phase transitions in the phase coexistence region
such as droplet nucleation and annihilation or droplet-strip transitions
for systems with periodic boundary conditions.
We characterize the efficiency of the optimized histogram sampling by measuring round-trip
times $\tau(N,Q)$ across the phase transition for samples of size $N$ spins. While we find power-law scaling of $\tau$ vs. $N$ for small $Q \lesssim 50$ and $N \lesssim 40^2$, we observe a crossover to exponential scaling for larger $Q$.
These results demonstrate that despite the ensemble optimization broad-histogram simulations
cannot fully eliminate the supercritical slowing down at strongly first-order transitions.
\end{abstract}

\pacs{02.70.Rr,  64.60.De, 75.10.Hk}
% 02.70.Rr General statistical methods
% 64.60.De Statistical mechanics of model systems (Ising model, Potts model, field-theory models, Monte Carlo techniques, etc)
% 75.10.Hk Classical spin models

\maketitle

%%%%%%%%%%%%%%%%%%%%%%%%%%%%%%%%%%%%%%%%%%%%%%%%%%%
% Introduction ---
%%%%%%%%%%%%%%%%%%%%%%%%%%%%%%%%%%%%%%%%%%%%%%%%%%%

\section{Introduction}
\label{sec:introduction}

Competing phases or interactions in many-particle systems can give rise to complex free energy landscapes that exhibit multiple local minima and maxima. Thermal equilibration in these systems slows down exceedingly due to the (exponential) suppression of tunneling across free energy barriers.
Examples of such slowly equilibrating systems can be found in various settings ranging from spin glasses %over frustrated magnets 
to folded proteins.

Numerical approaches to simulate these systems in thermal equilibrium suffer from the same slowing down %and a proliferation of autocorrelation times 
when based on variational techniques or conventional Monte Carlo sampling. To overcome this bottleneck various alternative sampling approaches have been developed in recent years.
Most of these approaches, which include multicanonical sampling~\cite{berg1991,berg1992}, broad-histogram sampling~\cite{oliveira1996}, parallel tempering~\cite{hukushima1996,marinari1992, lyubartsev1992}, multiple Gaussian modified ensemble~\cite{neuhaus2007}, and Wang-Landau sampling~\cite{wang2001, wang2001-1}, can be broadly grouped as extended ensemble techniques. Their common goal is to sample a statistical ensemble that allows to significantly broaden the range of sampled energies beyond the comparatively narrow distribution of the canonical ensemble.

The Wang-Landau algorithm tries to bring the idea of a broad sampling to an extreme by sampling a flat histogram in energy space. However, it was soon realized that sampling a uniform energy distribution is not necessarily the optimal way to improve equilibration and reduce autocorrelation times~\cite{neuhaus2003,dayal2004,alder2004}.
Instead it turns out that in order to (considerably) speed up equilibration and minimize autocorrelation times one should sample a {\em non-uniform} energy distribution that allocates more statistical weight to the bottleneck(s) of the simulation which typically coincide with the free energy barriers~\cite{trebst2004}. These so-called optimized ensembles are tailored to a given physical system and directly reflect the underlying free energy landscape.
One can systematically obtain the optimized statistical ensemble from an initial broad-histogram distribution by applying a feedback algorithm~\cite{trebst2004} that reallocates statistical weight based on measurements of the (local) diffusivity of the random walk which the system performs in energy space during the simulation.
This ensemble optimization has been applied in a broad variety of physical systems suffering from long
thermal equilibration times in the absence of efficient non-local updates including folded proteins~\cite{trebst2006,escobedo2008,binder2008,christen2008}, frustrated magnetic systems~\cite{bergmann2007,chern2008},  and dense liquids~\cite{trebst2005}.
The technique has further been generalized to optimize the grid of temperature points used in parallel tempering
simulations~\cite{katzgraber2006}, has been used in combination with cluster udpates~\cite{yu2005} and has been adopted for the simulation of quantum systems~\cite{wessel2007,harada2008}.

\begin{figure}[t]
%	\centering
	\includegraphics[width=\columnwidth]{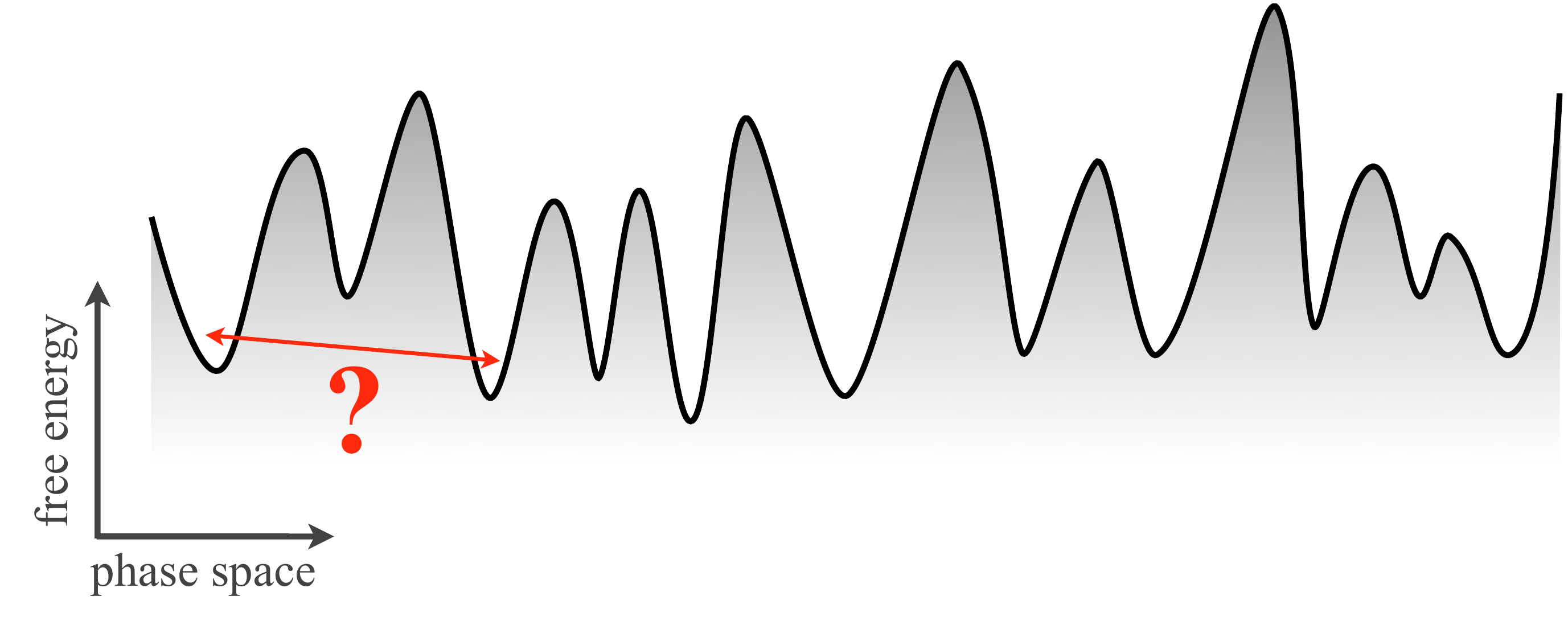}
	\caption{(color online)
	Sketch of the free energy landscape in phase space for a slowly equilibrating system.}
	\label{fig:landscape}
\end{figure}

In this manuscript, we apply and analyze the ensemble optimization technique in the context of a strong first-order phase transition where the characteristic double-well structure of the free energy provides a generic situation for entropically suppressed equilibration. In particular, we consider the thermal phase transition of the $Q$-state Potts model in the limit of large $Q$, with our calculations being performed for up to $Q=250$ states.
We find that the optimized ensemble aims to overcome the entropic barrier(s) of this transition by allocating most of the statistical weight in the energy range that corresponds to phase coexistence, e.g. the suppressed energy region of the characteristic bimodal distribution of the canonical ensemble. Remarkably, a multi-peak structure evolves in the optimized histogram that clearly resolves various intermediate transitions between metastable states, such as droplet formation and droplet-strip transitions.

The remainder of the manuscript is structured as follows: We will first provide a brief review of the ensemble optimization technique in Section~\ref{sec:ensembles}. In the subsequent Section~\ref{sec:potts} we turn to the large-$Q$ Potts model and discuss the multiple distinct features of the optimized broad-histogram distribution. We conclude our analysis by measuring the performance of the optimized ensemble technique in Section~\ref{sec:performance}.

%%%%%%%%%%%%%%%%%%%%%%%%%%%%%%%%%%%%%%%%%%%%%%%%%%%
% Introduction ---
%%%%%%%%%%%%%%%%%%%%%%%%%%%%%%%%%%%%%%%%%%%%%%%%%%%

\section{Optimized Ensembles}
\label{sec:ensembles}

We start our discussion of the ensemble optimization technique by first offering a broader view on Monte Carlo sampling and statistical ensembles. We then briefly review the derivation of the optimized ensembles and a related feedback algorithm.

\subsection{Monte Carlo sampling and statistical ensembles}
Speaking in broader terms one might take the perspective that the idea underlying Monte Carlo sampling is to map a random walk in some high-dimensional space of configurations $\{ c_i \}$
\[
   c_1 \to c_2 \to \ldots \to c_i \to c_{i+1} \to \ldots
\]
onto a random walk in a lower dimensional space, such as energy space (which is a one-dimensional space)
\[
   E(c_1) \to E(c_2) \to \ldots \to E(c_i) \to E(c_{i+1}) \to \ldots
\]
and to define a statistical ensemble in this latter low-dimensional space which then determines the transition probabilities between configurations in the original high-dimensional space. In particular, the statistical ensemble assigns a statistical weight $w(c)$ to a configuration $c$ solely on the basis of the respective energy $E(c)$ of that configuration
\[
   w(c) \equiv w( E(c) ) \,.
\]
The most commonly used statistical ensemble, of course, is the canonical ensemble, where the statistical weights are defined as
\[
   w(c) = \exp{(-\beta E(c))} \,.
\]
In order to simulate a reversible Markov process in configuration space one then defines transition probabilities from configuration $c$ to $\tilde{c}$ such that detailed balance is fulfilled. Common choices for these transition probabilities are Metropolis weights
\[
   p_{\rm Metropolis}( c \to \tilde{c} ) = \min \left( 1, \frac{w(\tilde{c})}{w(c)} \right) \equiv \min \left( 1, \frac{w(E(\tilde{c}))}{w(E(c))} \right) \,,
\]
or heat-bath weights
\[
   p_{\rm heat-bath}( c \to \tilde{c} ) = \frac{w(\tilde{c})}{w(\tilde{c})+w(c)} \equiv  \frac{w(E(\tilde{c}))}{w(E(\tilde{c}))+w(E(c))}  \,.
\]
While these choices of transition probabilities indeed ensure that the random walk in configuration
space is Markovian, it should be noted that the projected random walk in energy space is {\em not} Markovian. This becomes clear when considering that multiple configurations may have the same energy $E$, whereas the distribution of energies that can be reached by a single update may be completely different for each of these configurations. Thus, there is additional information encoded in configuration space which is not captured by $E$, and it is this `memory' which makes the projected random walk in energy space non-Markovian.

\subsection{Non-uniform diffusivity and optimized histograms}

The random walk in energy space has another distinct feature: the (local) diffusivity of this random walk, which for a given energy level measures the ability of the random walker to move to other energy levels, is {\em not} uniform in energy space. In fact, it is exactly this modulation of the local diffusivity which reflects the roughness of the underlying energy landscape. A suppressed diffusivity signals a `bottleneck' of the simulation and is typically associated with a phase transition or other entropic barrier.

This modulation of the local diffusivity thus %offers a stark 
differentiates the various energy regimes for a given system, and in this light it becomes clear that %the principal 
one shortcoming of flat-histogram techniques is %exactly the 
that they use a uniform distribution of statistical weight across these inherently different energy regimes.
In contrast the optimized ensemble method allocates statistical weight based on measurements of the local diffusivity and shifts additional statistical weight towards the bottleneck(s) of the simulation, e.g. those energy regimes with a suppressed local diffusivity. As a result the so-optimized random walk in energy space will sample a non-uniform histogram, spend more time in energy regimes with low diffusivity, and thereby do its best to suppress the bottlenecks associated with the underlying free energy landscape.

In more technical terms, we consider a random walk in some energy range $[E_-, E_+]$ between two extremal energies $E_-$ and $E_+$.
In this paper we sample the entire energy range, so $E_-$ and $E_+$ are, respectively, the lowest and highest energies that our model has.
The random walkers in energy space will drift between these two extremal energies and we can think of the overall random walk as being composed of two opposite steady-state `currents' between these two extremal energies. These two currents exactly compensate one another, as the system remains in equilibrium, and are independent of energy.
We can express these currents as
\begin{equation}
  j = D(E) H(E) \frac{df}{dE} \,,
  \label{eq:current}
\end{equation}
where $D(E)$ is the local diffusivity in energy space, $H(E)$ is the sampled energy histogram and $f(E)$ defines the orientation of the current by measuring for a given energy the fraction of random walkers which have last visited one of the two extremal energies, say, the lower extremal energy $E_-$. This latter fraction can be measured by recording two histograms, $H_+(E)$ and $H_-(E)$, where, for each Monte Carlo step, one increments the histogram with label `+' or `-' depending on which extremal energy the random walker has visited last. The two histograms $H_+(E)$ and $H_-(E)$ thus sum up to the total histogram $H(E) = H_+(E) + H_-(E)$. The fraction $f(E)$ is then given by $f(E) = H_-(E)/H(E)$.

In order to speed up equilibration one wants to maximize the current~\eqref{eq:current} between the two extremal energies. Varying the histogram $H(E)$ this can be achieved~\cite{trebst2004} by sampling a non-uniform distribution
\begin{equation}
H^{\text{opt}}(E) \propto \frac{1}{\sqrt{D(E)}}
\label{eq:OptimizedHistogram}
\end{equation}
that is inversely proportional to the square root of the local diffusivity $D(E)$ and thus reallocates statistical weight to those energy levels with suppressed diffusivity.

\subsection{The feedback algorithm}

In order to sample the optimized ensemble~(\ref{eq:OptimizedHistogram}) we apply the feedback algorithm outlined in Ref.~\cite{trebst2004}. We start from an initial broad-histogram ensemble with statistical weights $w(E)$ which we obtain from a few iterations of the Wang-Landau algorithm
or by extrapolating results from smaller system sizes. Running a (short) simulation for this initial ensemble we record the two histograms $H_+(E)$ and $H_-(E)$ introduced above which in turn allow us to calculate the local diffusivity as
\begin{equation}
    D(E) \propto \left( H(E) \frac{df}{dE} \right)^{-1} \,.
    \label{eq:diffusivity}
\end{equation}
We then refine the statistical weights by feeding back this local diffusivity and define new optimized weights as
\begin{equation}
   w_{\rm opt}(E) = w(E)\sqrt{ \frac{1}{H(E)} \frac{df}{dE} } \,.
\end{equation}
Subsequent simulations are performed for this new set of statistical weights. To further improve and eventually converge the statistical weights for the optimized ensemble we repeat the feedback procedure several times. Note that in order to ensure convergence the number of Monte Carlo steps between subsequent feedback iterations needs to be increased; we typically double the number of Monte Carlo steps for consecutive runs.

\subsection{Improving the first feedback step}

There is a certain trade-off in performing the early feedback steps in the algorithm outlined above:
On the one hand, an early feedback after only a small number of Monte Carlo sweeps appears advantageous as it may quickly give dramatically improved statistical weights and thereby speed up all subsequent simulations. On the other hand, the quality of the feedback is rather sensitive to noisy input data, especially in calculating the numerical derivative $df/dE$ used in the feedback.
To minimize this trade-off one thus needs a way to quickly estimate this latter derivative in the presence of (substantial) noise. Conventional approaches such as finite-difference formulas, however, turn out to be exquisitely sensitive  to the noise in the recorded histograms $H_+(E)$ and $H_-(E)$. In particular, the measured fraction $f(E) = H_-(E)/H(E)$ is a monotonically decaying function only when the simulation is in equilibrium in the simulated statistical ensemble which for a suboptimal choice, such as the flat-histogram ensemble, may require rather long Monte Carlo runs.

We have therefore developed a scheme that allows for the estimation of the derivative in the presence of significant noise. The idea is to analyze the measured fraction $f(E)$ in Fourier space, truncate the high-frequency terms which can be associated with noise, and then determine the derivative using the low-frequency terms only. In doing so, we make use of the fact that for a continuous Fourier transformation
\[
\tilde{f}(\omega) = \frac{1}{\sqrt{2\pi}} \int_{-\infty}^{\infty} e^{-i \omega E} f(E) dE \,,
\]
the derivative of the original function $f(E)$
\begin{equation}
\tilde{\partial}_E f(E) = \frac{1}{\sqrt{2\pi}} \int_{-\infty}^{\infty} i\omega \cdot e^{i\omega E}\ \tilde{f}(\omega)\ d\omega
\end{equation}
can be easily calculated in Fourier space and then transformed back.

In implementing this idea one needs to work around several obstacles. First, in order to avoid irrelevant boundary terms, the function to be analyzed using the Fourier transformation should be periodic. We therefore concatenate $f(E)$ with its reflection.
Secondly, the above relation strictly holds only for the continuous Fourier transformation. As the energy levels of the Potts model and, in general, the energy bins of a broad histogram simulation are discrete, we need to work with a discrete Fourier transformation. To overcome errors introduced by this, we make use of the following iterative scheme which refines the calculated derivative by iteratively reducing the deviation between the integral of the approximated derivative and the original function:
\begin{eqnarray*}
\delta f^1 &=& \tilde{\partial}_E f(E) \\
\delta f^2 &=& \delta f^1 + \tilde{\partial}_E \left(f(E) - \int \delta f^1 dE \right) \\
&\ldots& \\
\delta f^{i+1} &=& \delta f^{i} + \tilde{\partial}_E \left(f(E) - \int \delta f^{i} dE \right) \\
\end{eqnarray*}
Here, $\tilde{\partial}_E$ denotes the approximate derivative using the Fourier-based scheme above. The scheme is iterated until the norm of the correction term falls below a certain threshold.

% Updates
%A significant reduction of computation time, in particular for large values of $Q$, can be obtained by using heat bath instead of Metropolis updates. For heat bath updates, single-site flips are not proposed with equal probability, but instead with a probability proportional to the weight of the proposed configuration. The chosen flip is accepted with probability 1. The increased time necessary to calculate the weights for one flip is more than outweighed by the improved dynamics.

%%%%%%%%%%%%%%%%%%%%%%%%%%%%%%%%%%%%%%%%%%%%%%%%%%%
% The large-Q Potts model ---
%%%%%%%%%%%%%%%%%%%%%%%%%%%%%%%%%%%%%%%%%%%%%%%%%%%

\section{The large-Q Potts model}
\label{sec:potts}

The two-dimensional $Q$-state Potts model is well known~\cite{wu1982}
to undergo a thermal phase transition
which turns from continuous for small $Q \leq 4$ to weakly first-order for $Q > 4$ and
eventually becomes a strong first-order transition for $Q \gg 5$.
We will turn to this latter case of a strong first-order transition in systems with up to $Q=250$
different Potts states to explore the extent to which the optimized ensemble algorithm can achieve
equilibration at such a transition.

The Hamiltonian of the $Q$-state Potts model is given in terms of  spins $\sigma_i$ which
take discrete values $\sigma_i \in \lbrace 1,\ldots,Q \rbrace$ as
\begin{equation}
H = -\sum_{\langle i,j \rangle} \delta \left({\sigma_i, \sigma_j} \right) \,.
\end{equation}
Here the sum runs over all pairs of nearest neighbors on a square lattice, and the Kronecker $\delta$-function
tests whether two Potts spins have the same values. %point along the same direction.
We have run simulations for two sample geometries: %types of boundary conditions: 
a ``toroidal geometry", i.e. a square lattice with periodic boundary conditions, and a ``cube geometry" by forming a cube %out of six 
with square lattices on each of its 6 faces.
%All our simulations are performed for periodic boundary conditions.

We will start our discussion by briefly mentioning both exact and
numerical results for thermodynamic properties of the Potts model in this large, but finite $Q$ limit.
We will then turn to the energy regime associated with phase coexistence at this first-order phase transition and examine the various intermediate, metastable states such as droplets or strips which occur in this regime.
In particular, we will discuss a distinct multi-peak structure which emerges in the optimized histogram distribution and show how these features can be linked to transitions between the various metastable states.

\begin{figure}[htp]
	\includegraphics[width=\columnwidth]{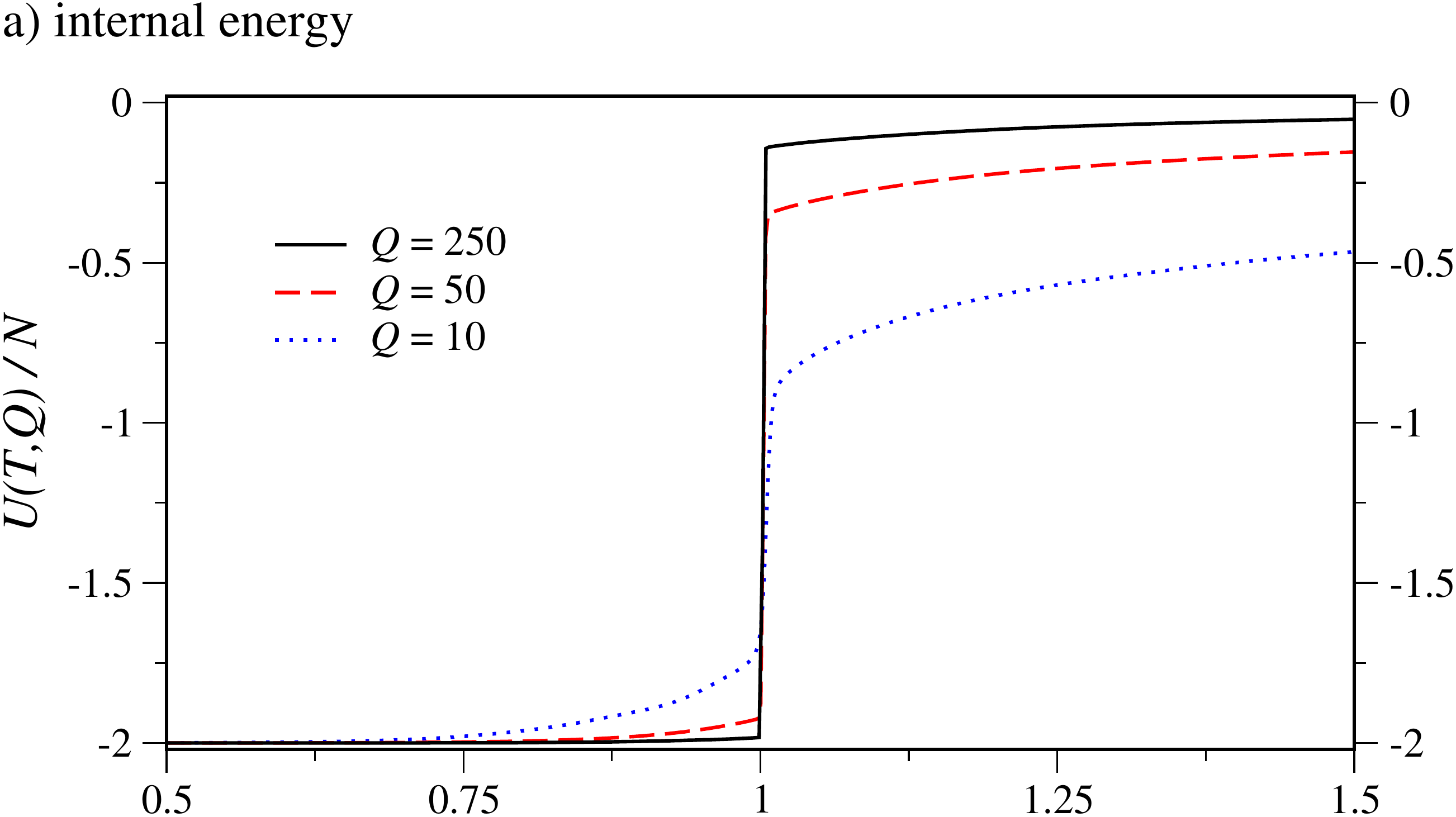}
	\vskip 3mm
	\includegraphics[width=\columnwidth]{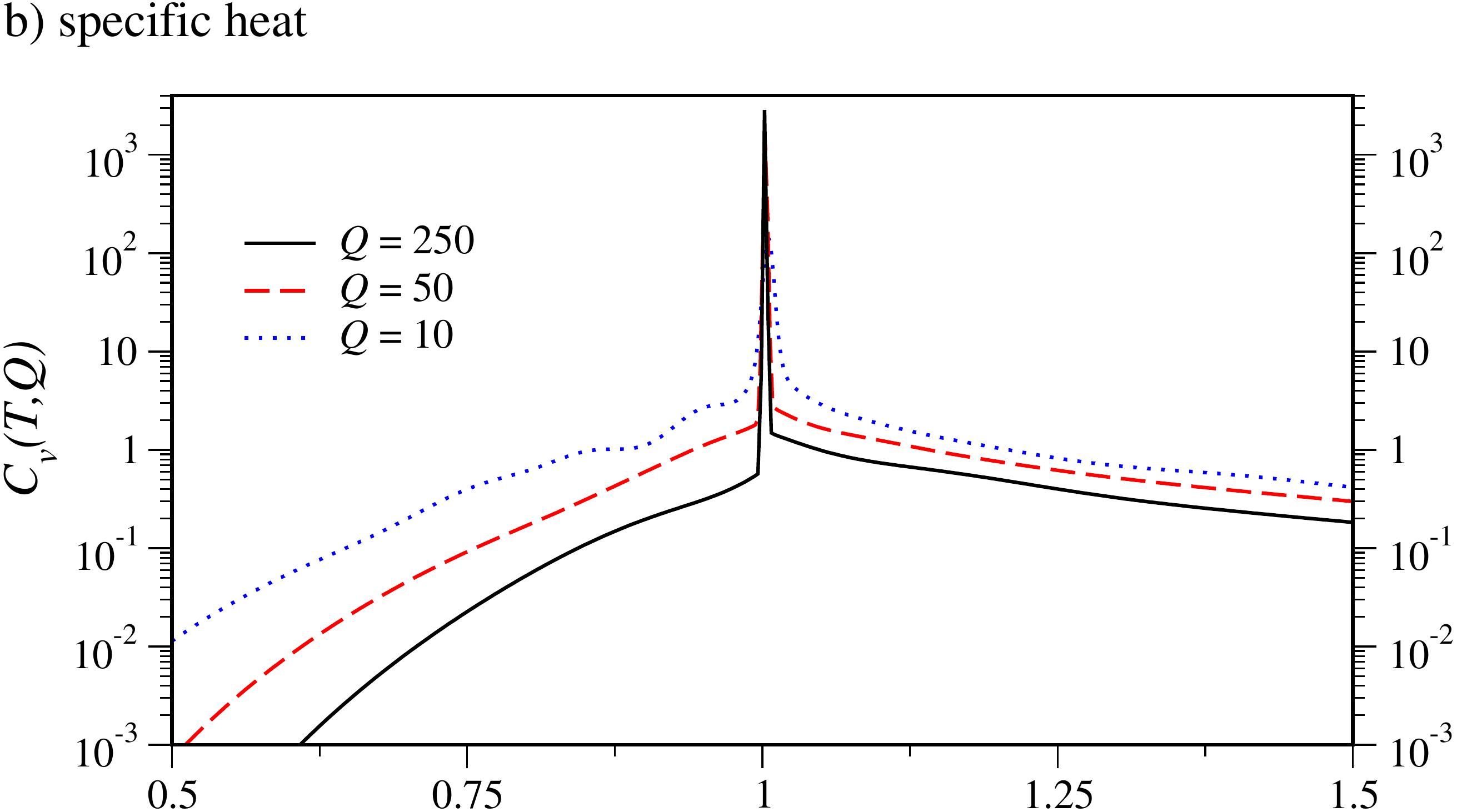}
	\vskip 3mm
   	\includegraphics[width=\columnwidth]{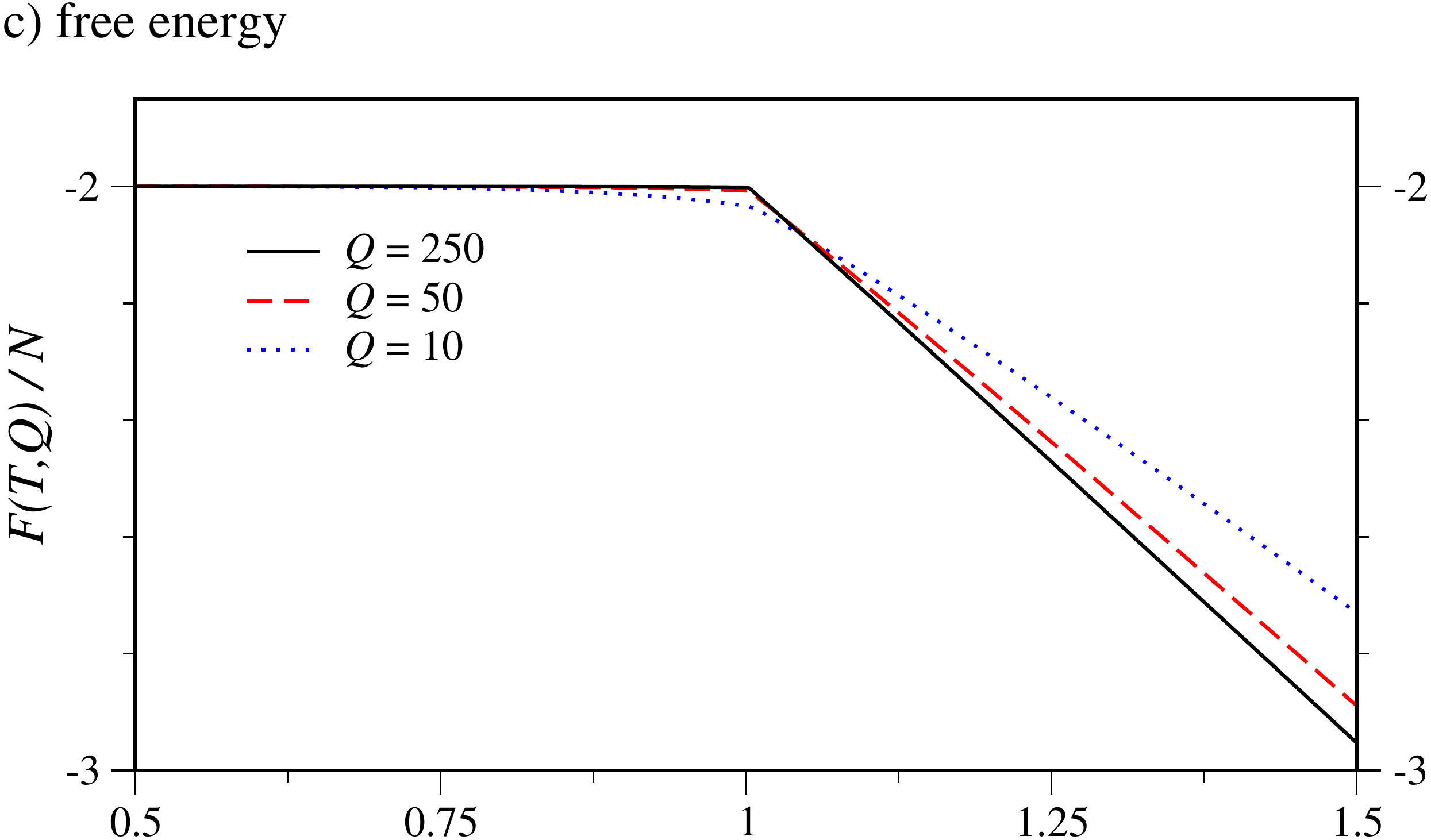}
 	\vskip 3mm
	\includegraphics[width=\columnwidth]{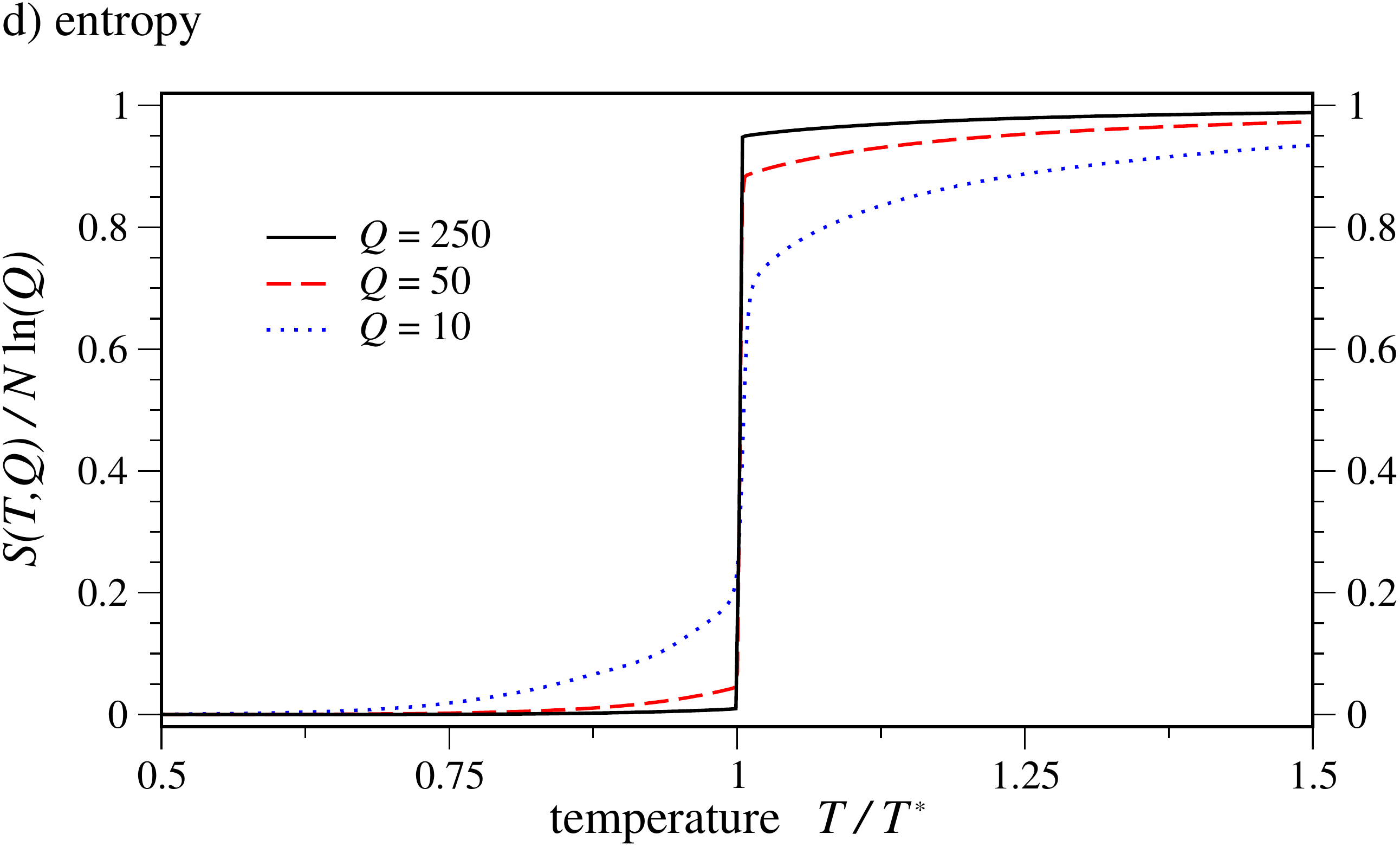}
	\caption{(color online)
	               Thermodynamic properties of the $Q$-state Potts model in the large, but finite $Q$ limit:
	               a) energy,
	               b) specific heat,
	               c) free energy, and
	               d) entropy.
	               The temperature axis is rescaled by the transition temperature
	               $T^* = 1/\ln{(1+\sqrt{Q})}$.
	               Data shown is for system size $L=22$ with periodic boundary conditions.}
	\label{fig:therm}
\end{figure}

%%%%%%%%%%%%%%%%%%%%%%%%%%%%%%%%%%%%%%%%%%%%%%%%%%%
\subsection{Thermodynamic properties}

\begin{figure*}[t]
\begin{tabular}{ccccccc}
	a) $\ E/2N = -0.20$ & \hspace{8mm} & b) $\ E/2N = -0.35$ & \hspace{8mm} &
	c) $\ E/2N = -0.54$ & \hspace{8mm} & d) $\ E/2N = -0.87$ \vspace{2mm} \\
	\includegraphics[width=1.2in]{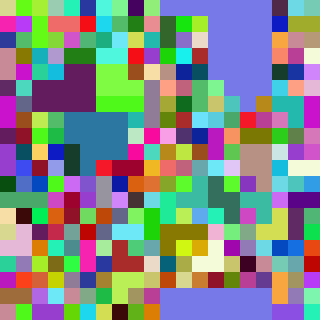} & \ &
	\includegraphics[width=1.2in]{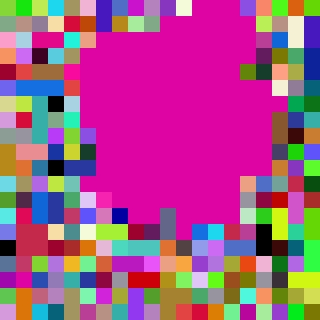} & \ &
	\includegraphics[width=1.2in]{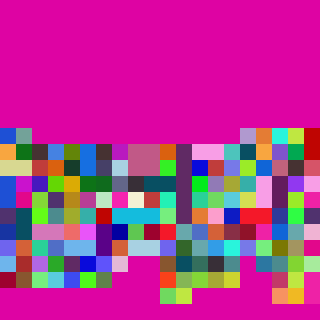} & \ &
	\includegraphics[width=1.2in]{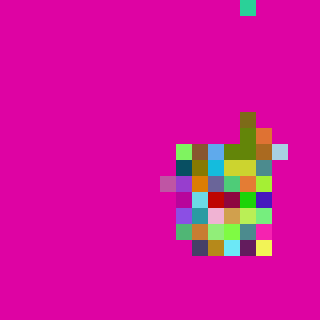}
\end{tabular}
	\caption{(color online)
	                Snapshots of spin configurations in the phase coexistence region of
	                the 250-state Potts model.
	                a) Formation of several small ordered droplets within the disordered phase.
	                b) A dominant droplet is formed.
	                c) The droplet percolates to a strip wrapping around one boundary.
	                d) A single disordered droplet remains in the ordered phase.
	                All data shown is for system size $L=20$.}
	\label{fig:metastable}
\end{figure*}

The thermal phase transition in the $Q$-state Potts model from a disordered phase
at high temperatures to an ordered phase at low temperatures occurs at a transition temperature
$T^*$ which for the infinite system is found~\cite{wu1982} to be
\begin{equation}
T^* = \frac{1}{\ln(1+\sqrt{Q})} \,.
\end{equation}
This phase transition is accompanied by  sharp features in various thermodynamic properties
such as the energy, the specific heat, the free energy and the entropy.
Since the optimized ensemble algorithm allows to directly calculate the density of
states $g(E)$, we can readily compute  all of these thermodynamic variables.
Our results are summarized in Fig.~\ref{fig:therm} for simulations
with $Q=10,50,250$ states, where we have rescaled the temperature axis by the %with regard to the
transition temperature $T^*$ %of the $Q$-state Potts model
in the thermodynamic limit.
As expected, the features associated with the phase transition sharpen
as we increase the number of Potts states $Q$ for a system of fixed size $L$. For instance, the discontinuous jump
in the energy grows with increasing $Q$ and $U(T)$ approaches a step function
in the limit of $Q \to \infty$. It is this broadening energy regime within the discontinuous jump
of the energy that is associated with phase coexistence and the occurrence
of intermediate, metastable states as we will discuss in detail below.

%The collapse of the entropy for $Q=250$ shows how the infinite-$Q$ limit is approached: the asymptotic value of the entropy, $S(T \rightarrow \infty) = \ln (Q)$, which is also the slope of the free energy for $T > T_c, T \rightarrow \infty$, is approached very quickly.

%%%%%%%%%%%%%%%%%%%%%%%%%%%%%%%%%%%%%%%%%%%%%%%%%%%
\subsection{Phase coexistence and metastable states}

The distinct characteristic of a first-order phase transition is a free energy profile
that passes through a double well shape as one drives the transition with some
external parameter such as temperature. At the transition temperature the two
minima of the free energy are exactly equal leading to coexistence of the two
phases.
For the high-$Q$ Potts model at hand there is a considerable amount of latent heat
associated with this transition, i.e. the internal energies of the two phases in
proximity to this phase transition vastly differ
%as becomes apparent in the discontinuous jump of the internal energy
as shown in Fig.~\ref{fig:therm}a).
As the system goes from one phase to the other this latent heat
is not released (or absorbed) in a single step, but the system undergoes a
{\em sequence} of phase transitions between various  {\em metastable} states
which are not minima of the free energy in thermal equilibrium, but correspond
to states with intermediate internal energies.
One such metastable state is a droplet of one phase inside the other phase.  Since
the free energy density of the two phases becomes arbitrarily close in the vicinity of the
transition, the free energy cost of forming a droplet is due to the surface
of the droplet, and not to its volume. It is thus %energetically
entropically favorable to nucleate and
grow a single droplet of a shape that minimizes its surface free energy.
This droplet condensation transition has recently been studied in detail for a variety
of physical systems using both
numerical~\cite{neuhaus2003,macdowell2006,nussbaumer2006,nussbaumer2008}
and analytical~\cite{biskup2002,biskup2003,macdowell2004,binder2003,biskup2004}
approaches.

For a torus geometry, e.g. a %finite two-dimensional
system with periodic boundary conditions,
this droplet will subsequently expand as the total energy is changed until it
percolates and it becomes entropically more favorable to form a strip wrapping around
one of the boundaries.
As this strip further grows the role of the two phases will eventually be reversed and the
system will undergo a second sequence going from a strip to a droplet and eventually
annihilate the remaining droplet to complete the phase transition from one phase to the
other. This transition was first discussed for the Ising model in an external magnetic field
by Leung and Zia~\cite{leung1990} and studied in detail by Neuhaus and Hager~\cite{neuhaus2003} using multicanonical Monte Carlo sampling.
The droplet nucleation and droplet-strip transitions were also observed for the Potts model
with $Q=10$ and system sizes of up to $1024 \times 1024$ spins using a microcanonical
approach~\cite{mayor2007}.

We show representative snapshots of spin configurations reflecting these metastable
states in Fig.~\ref{fig:metastable}. All snapshots have been taken taken from our numerical
simulations of the 250-state Potts model.

\begin{figure}[b]
	\includegraphics[width=\columnwidth]{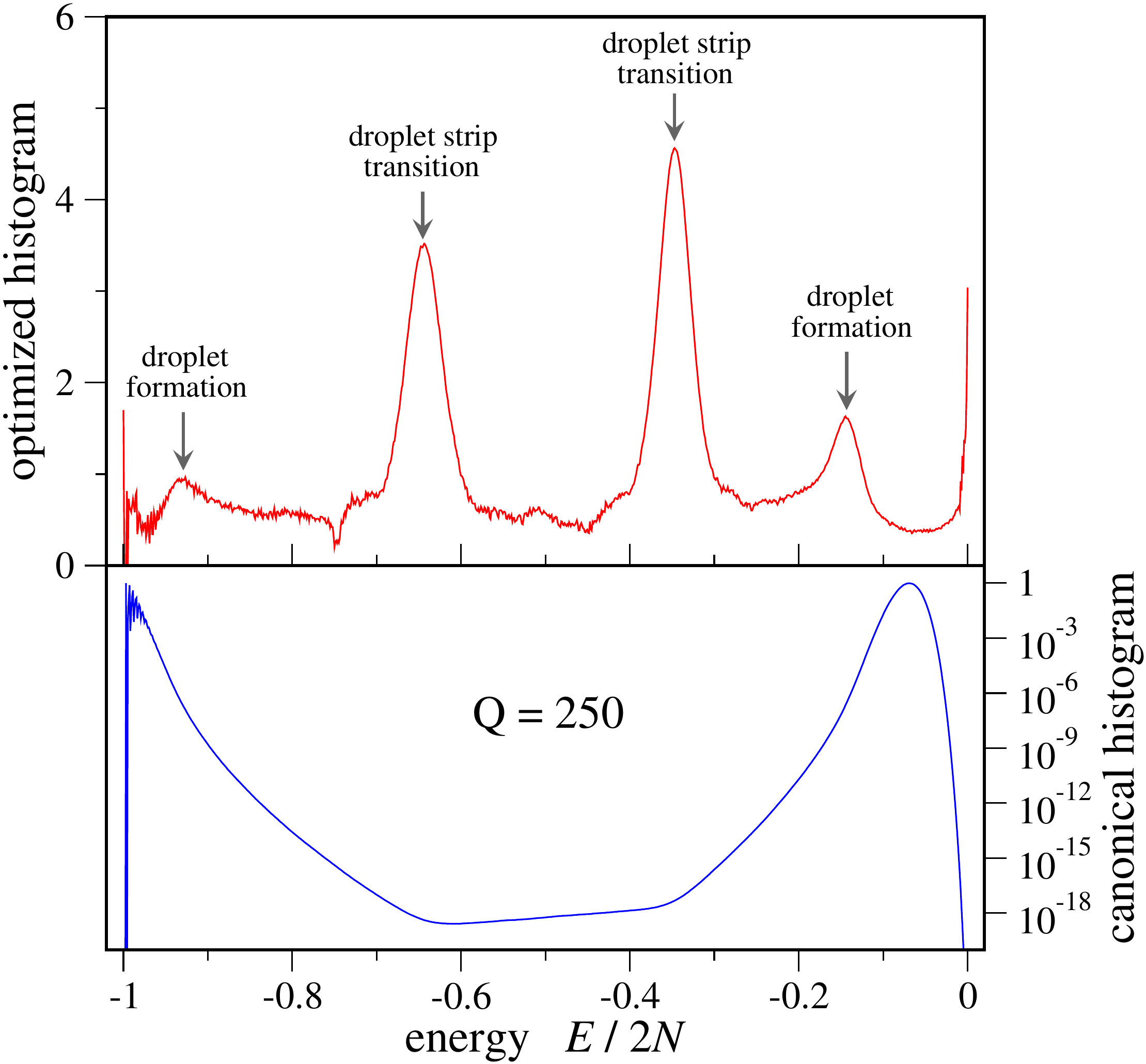}
	\caption{(color online)
	Histograms of the optimized ensemble (top panel) and the
	canonical ensemble at the transition temperature $T^*$ (bottom panel)
	for the 250-state Potts model with $L=22$ and toroidal geometry.
	In contrast to the bimodal distribution of the canonical ensemble
	the histogram of the optimized ensemble reveals a distinct four-peak structure
	reflecting the transitions between the various metastable states in the phase
	coexistence region.}
	\label{fig:canon}
\end{figure}

\subsection{Droplet nucleation and droplet-strip transition}

For the toroidal system we can thus distinguish four intermediate transitions taking place ``within'' a first-order
transition: % for a two-dimensional system with periodic boundary conditions:
The nucleation of a dominant droplet (which might occur via the condensation of multiple
small droplets), a droplet to strip transition and two more processes where the roles of the
two phases are reversed.
%We will find that each of these transitions is associated with
%an entropic barrier that is not fully suppressed in our feedback-optimized ensemble.
%causes the long lifetimes of the intermediate metastable states.
%Another way to see this is that t
These intermediate transitions occur at energies that are within the discontinuous
jump of the internal energy $U(T)$ plotted in Fig.~\ref{fig:metastable}~a) and are therefore not
equilibrium energies at any temperature.  %As a consequence, t
For the canonical ensemble the states at these intermediate transitions are strongly suppressed,
%which in turn gives rise to the
with its characteristic bimodal distribution of
sampled energies as shown in the bottom panel of Fig.~\ref{fig:canon}.

\subsubsection{Multi-peak structure}

%To equilibrate rapidly across a strongly first-order phase transition %overcome the bottleneck of the canonical ensemble
%we now turn to the extended ensemble
%approach outlined in Sec.~\ref{sec:ensembles} that aims at optimally sampling a broad histogram in
%energy space.

In sharp contrast to the canonical ensemble the feedback algorithm of Sec.~\ref{sec:ensembles}
reallocates significant statistical weight to the energy range located {\em within} the double-peak
structure of the canonical distribution corresponding to the discontinuous jump of the internal energy.
Strikingly, we find the emergence of a distinct multi-peak structure in this energy
range as shown in the top panel of Fig.~\ref{fig:canon}.
The emergent peaks resolve precisely the four intermediate transitions discussed above.
We come to this identification, as given in Fig.~\ref{fig:canon}, by
i) comparing the energies of typical configuration snapshots as shown in Fig.~\ref{fig:metastable}
with the locations of these peaks,
ii) estimating the transition energies of the droplet-strip transitions as discussed in
Section~\ref{sec:location-droplet-strip}
and iii) calculating order parameters for the droplet-strip transitions as
detailed in Section~\ref{sec:order-parameter-droplet-strip}.
The redistribution of statistical weight in this multi-peak structure also reveals that these
transitions between metastable states are of different severity. With the histogram peaks
corresponding to the droplet-strip transitions being much more pronounced than those
corresponding to droplet nucleations we can conclude that the entropic barriers associated
with the former transitions are significantly larger than those associated with the latter transitions.
Another observation regarding this emerging multi-peak structure is that the histogram distribution
is not perfectly symmetric with respect to the ordered / disordered phases. For instance, the difference
of the two smaller peaks reflects that droplet formation in the disordered phase is associated
with a larger entropic barrier than droplet formation in the ordered phase.

\begin{figure}[t]
	\includegraphics[width= \columnwidth]{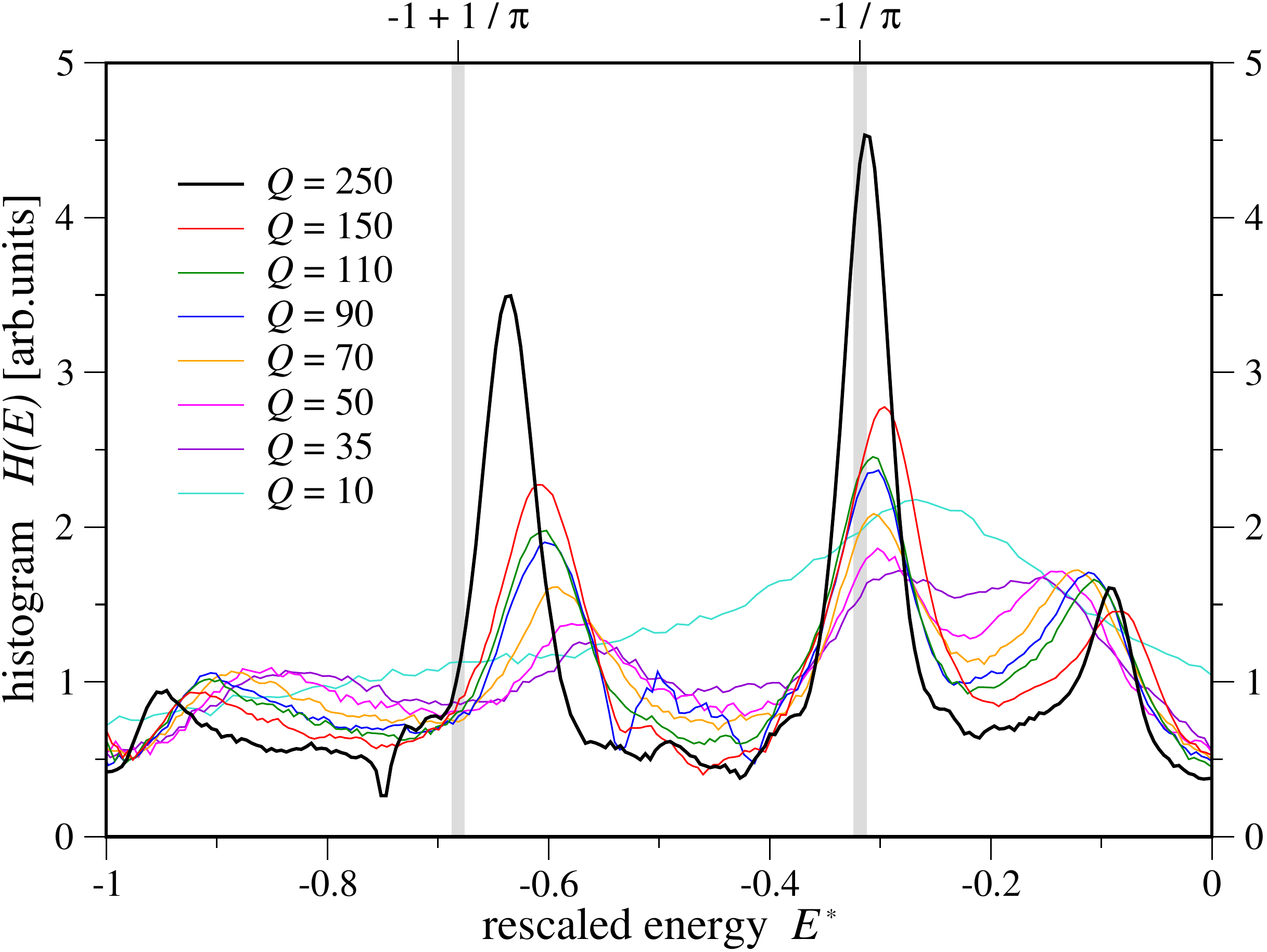}
	\caption{(color online)
	                Optimized histograms for increasing number of Potts states $Q$ and fixed system size
	                $L=22$ with toroidal geometry.
	                While the histogram peaks in the emerging multi-peak structure seem to diverge
	                with increasing $Q$ for intermediate energies and associated with the droplet-strip
	                transitions,
	                the histogram peaks associated with droplet formations appear to converge.
	                The plotted energy range corresponds to the coexistence region defined in
	                Eq.~\eqref{eqn:rescaling}.
	                }
	\label{fig:histQ}
\end{figure}

These characteristic features of the multi-peak structure further evolve as we vary the strength of
the underlying first-order transition by increasing the number of Potts states $Q$ or the system size $L$
as shown in Figs.~\ref{fig:histQ} and \ref{fig:histL}, respectively.
With increasing the number of Potts states $Q$ we find the droplet-strip transitions to attract
considerably more statistical weight than the droplet formation transitions. In particular, the histogram
peaks associated with the droplet-strip transitions seem to diverge with increasing $Q$, while the
histogram peaks associated with the droplet formation transitions appear to converge to a finite height while sharpening with increasing $Q$, see Fig.~\ref{fig:histQ}.
Similarly, we find that increasing the system size $L$ increases the peaks associated with the droplet-strip transitions more strongly than those associated with the formation of a droplet, as shown
in Fig.~\ref{fig:histL}. %this is indeed what we find.

%%%%%%%

\subsubsection{Location of droplet-strip transitions}
\label{sec:location-droplet-strip}

We can estimate the location of the intermediate droplet-strip transitions more quantitatively by
estimating the interface length of the droplet / strip on either side of the transition.
Making such an estimate for the droplet, however, requires knowledge of its rough shape.
The latter depends on the anisotropy of the surface tension and to some extent the geometry of the system.
% and can be determined using the Wulff construction\cite{wulff1901,laue1944,herring1951,burton1951}.
For the Potts model, the surface tension and consequently the equilibrium droplet shape have been calculated analytically for a droplet of fixed size embedded in an infinite volume~\cite{borgs1992,fujimoto1997}; for finite systems, Billoire {\it et al.}~\cite{billoire1994} have estimated the surface tension based on multi-canonical simulations of the canonical probability density for mixed-phase states.
% for a finite system it is unknown.
In the limit of large $Q$, which we consider here, the surface tension $\sigma$ is found to become isotropic and the droplet shape is expected to be roughly circular.
The location of the droplet-strip transition can then be estimated by identifying the radius of the droplet $R$ for which the surface free energy of the droplet, $F_{\text{droplet}} = 2 \pi \sigma R$, becomes equal to the surface free energy of the strip, $F_{\text{strip}} = 2 \sigma L$. The transition is therefore expected to take place at $R = L / \pi$.  At this transition point, the droplet occupies area $\pi R^2=L^2/\pi$ and thus a fraction $1/\pi$ of the total area of the sample.
%Under the assumption that all bonds across the interface of the two phases are broken and therefore do not contribute to the (inner) energy, w
We can estimate the total energy of a given domain pattern as a sum of the contributions from the domains and the interfaces.
For the example of an ordered droplet of radius $R$ in a disordered background we have
\begin{equation}
E = (L^2 - \pi R^2) e_{\text{dis}}(T) + \pi R^2 e_{\text{ord}}(T) + 2\pi Re_{\sigma}~.
\end{equation}
This configuration, by local stability of the curved interface, is at a temperature $T$ that is below the transition temperature $T^*$ by an amount
proportional to the curvature $1/R$ of the interface.  At that temperature, the energy densities of the two phases are
%where we determine the values
$e_{\text{ord}}(T)$ and $e_{\text{dis}}(T)$, while $e_{\sigma}$ is the excess energy per unit length in the interface.  % from the location of
%the peaks in the canonical histogram or, equivalently, from the step in the inner energy as we approach the transition temperature $\lim_{T \rightarrow T^*} U(T)$ from low/high temperatures, respectively.

\begin{figure}[t]
	\includegraphics[width= \columnwidth]{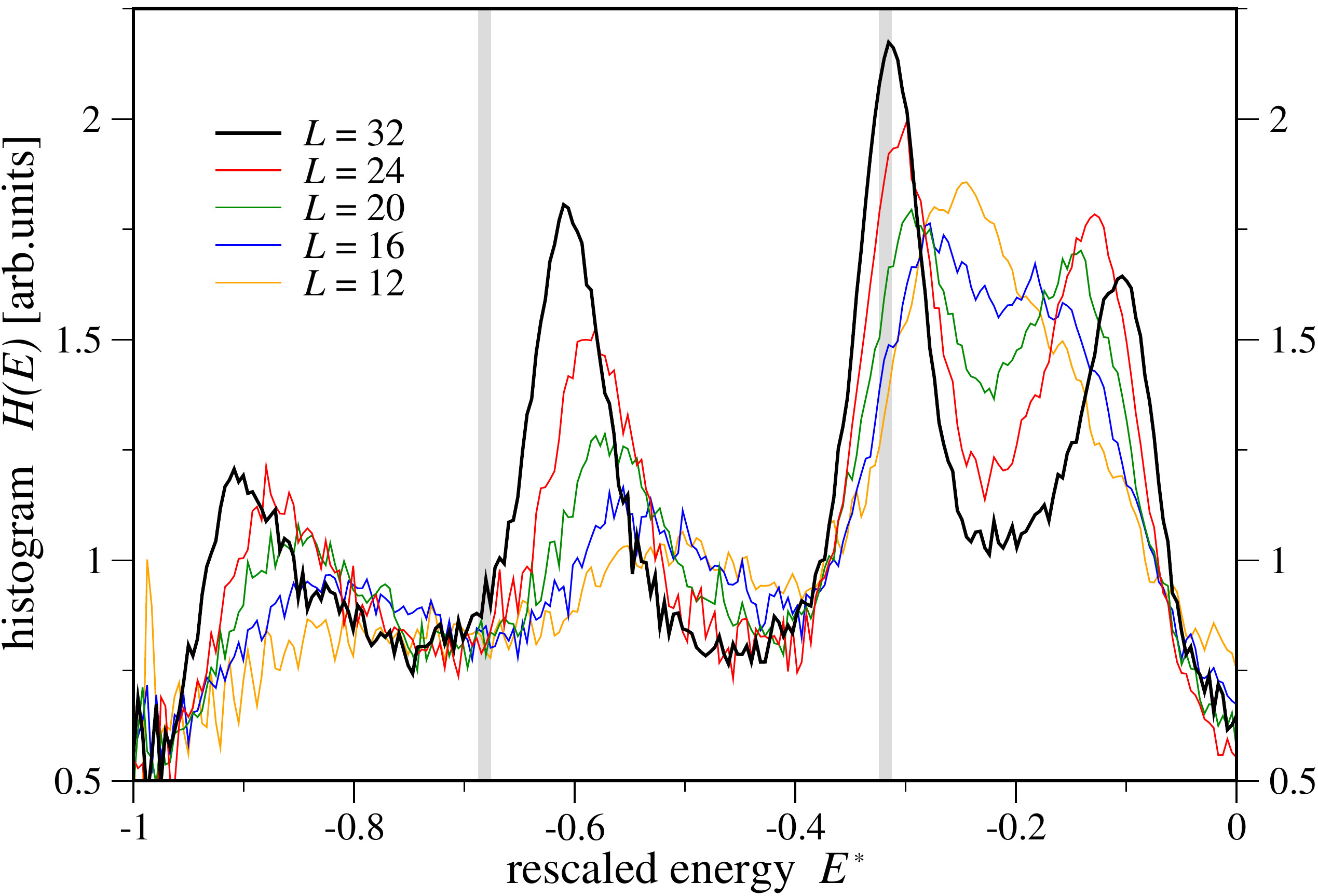}
	\caption{(color online)
	Optimized histogram for increasing system size $L$ and fixed number of Potts states $Q=50$.
	Similar to Fig.~\ref{fig:histQ} the peaks associated with the droplet-strip transitions proliferate
	more strongly with increasing system size than those associated with droplet formation.
         The plotted energy range corresponds to the coexistence region defined in
	Eq.~\eqref{eqn:rescaling}.
}
	\label{fig:histL}
\end{figure}

Simplifying this expression by keeping only the contributions that are
proportional to the areas of the domains, we obtain that the transition energies of the two droplet-strip transitions can be approximated as
\begin{eqnarray}
E^*_{\text{droplet-strip,1}} &=& - \frac{1}{\pi} \label{eqn:droplet-strip1} \,, \\
E^*_{\text{droplet-strip, 2}} &=& -1 + \frac{1}{\pi} \label{eqn:droplet-strip2} \,,
\end{eqnarray}
where these energies are given relative to the size of the coexistence region, e.g. the width of the jump of the internal energy plotted in Fig.~\ref{fig:metastable}a).
In the following, we rescale energies such that the energy of the ordered phase $E_{\text{ord}}(T^*)$ is mapped to $-1$ at the transition temperature $T^*$ and the energy of the disordered phase $E_{\text{dis}}(T^*)$ becomes $0$:
\begin{equation}
 E^* = \frac{E - E_{\text{ord}}(T^*)}{E_{\text{dis}}(T^*) - E_{\text{ord}}(T^*)} - 1.
 \label{eqn:rescaling}
\end{equation}
To rescale our numerical results we use the exactly-known energy densities of the two phases at
the transition temperature $T^*$ in the thermodynamic limit.

We have indicated the so-estimated locations of the two droplet-strip transitions by the vertical bars
in Figs.~\ref{fig:histQ}  and \ref{fig:histL}. Indeed, the respective histogram peaks associated with these transitions seem to converge to these locations in the limit of large $Q$ and $L$.
In more quantitative terms, the energy of the interface moves $E^*$ at the transitions to a higher energy by an amount proportional to $1/L$.  The small shift in $T$ due to the curvature of the interface of the droplet also moves $E^*$ at the transition by amount proportional (ignoring logs) to $1/(L\sqrt{Q})$ at large $Q$; this latter effect is of the same sign at the lower-energy droplet-strip transition, where the system is slightly ``superheated''.
The trends with increasing $Q$ and $L$ at this lower transition
can be clearly seen in Figs.~\ref{fig:histQ}  and \ref{fig:histL}, respectively.  At the higher-energy droplet-strip transition the two $~1/L$ finite-size effects are of opposite signs, so the peak in the histograms stays closer to $E^*=-1/\pi$.

At the energies of these droplet-strip transitions, the two configurations (droplet and strip) have the same entropy.  However for the system to make a transition between these two configurations, it must increase the amount of interface by an amount proportional to $L$.  The entropy deficit per unit length in the interface is proportional to $\log{Q}$ at large $Q$, so the entropy barriers at these transitions are proportional to $L\log{Q}$.

\subsubsection{Order parameter for the droplet-strip transition}
\label{sec:order-parameter-droplet-strip}

\begin{figure}[t]
	\includegraphics[width=\columnwidth]{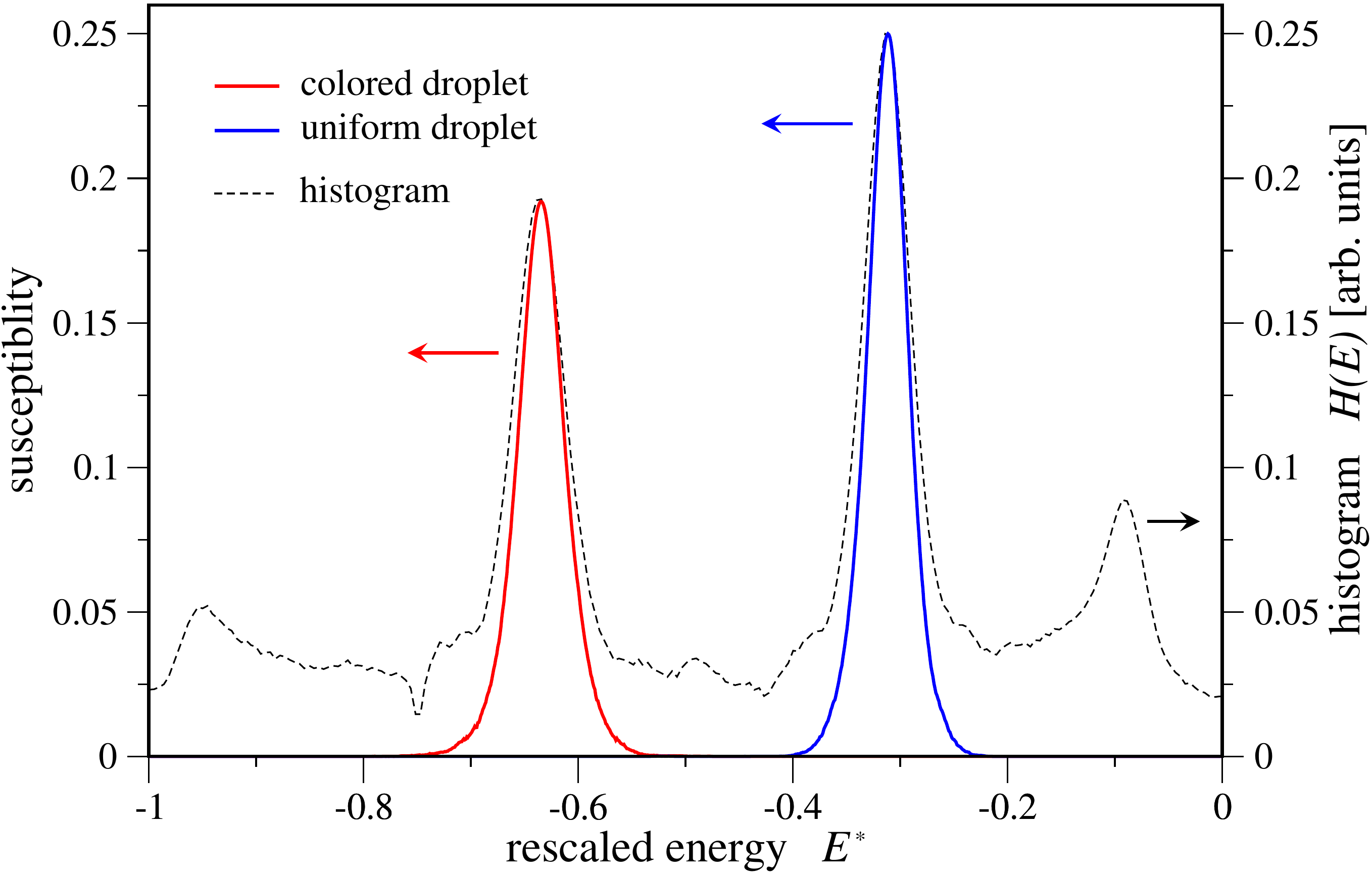}
	\caption{(color online)
	                Optimized histogram and susceptibilites for the droplet percolation order parameters.
	                The lower susceptibility peak has been rescaled to match the height of the peak in the
	                optimized histogram.
	                Data shown is for a 250-state Potts model and system size $L=24$.}
	 \label{fig:susc}
\end{figure}

An alternative approach to locate and further describe the droplet-strip transition is to define an order parameter for this transition. In doing so we follow an idea of Neuhaus and Hager~\cite{neuhaus2003} and measure the existence of a strip by measuring the dimensions $L_1, L_2$ of the minimal bounding box for the largest droplet in the system
\begin{equation}
O^{\text{dis}/\text{ord}} = \delta(L - \max(L_1^{\text{dis}/\text{ord}},L_2^{\text{dis}/\text{ord}})) \,,
\end{equation}
where the index "dis/ord" distinguishes whether the phase in the droplet corresponds to the disordered or ordered phase, respectively.
When the droplet percolates and a strip is formed, one dimension of the bounding box becomes equal to the system size $L$ and the order parameter jumps from 0 to 1. Furthermore, we can associate a susceptibility with this order parameter,
\begin{equation}
\chi_O = \langle O^2 \rangle - \langle O \rangle^2 = \langle O \rangle - \langle O \rangle^2 \in [0,1/4] \,,
\end{equation}
which we expect to proliferate at the droplet-strip transition. Comparing the divergence of this susceptibility with the respective intermediate peaks forming in the optimized histogram, we find that
they coincide not only in location, but also their respective shapes as shown in Fig.~\ref{fig:susc}.
The latter illustrates that the entropic barriers at this intermediate transition which the optimized ensemble overcomes by shifting statistical weight towards this transition arise solely from percolating a droplet into a strip.

\subsubsection{Droplet anisotropy}

\begin{figure}[t]
	\includegraphics[width=\columnwidth]{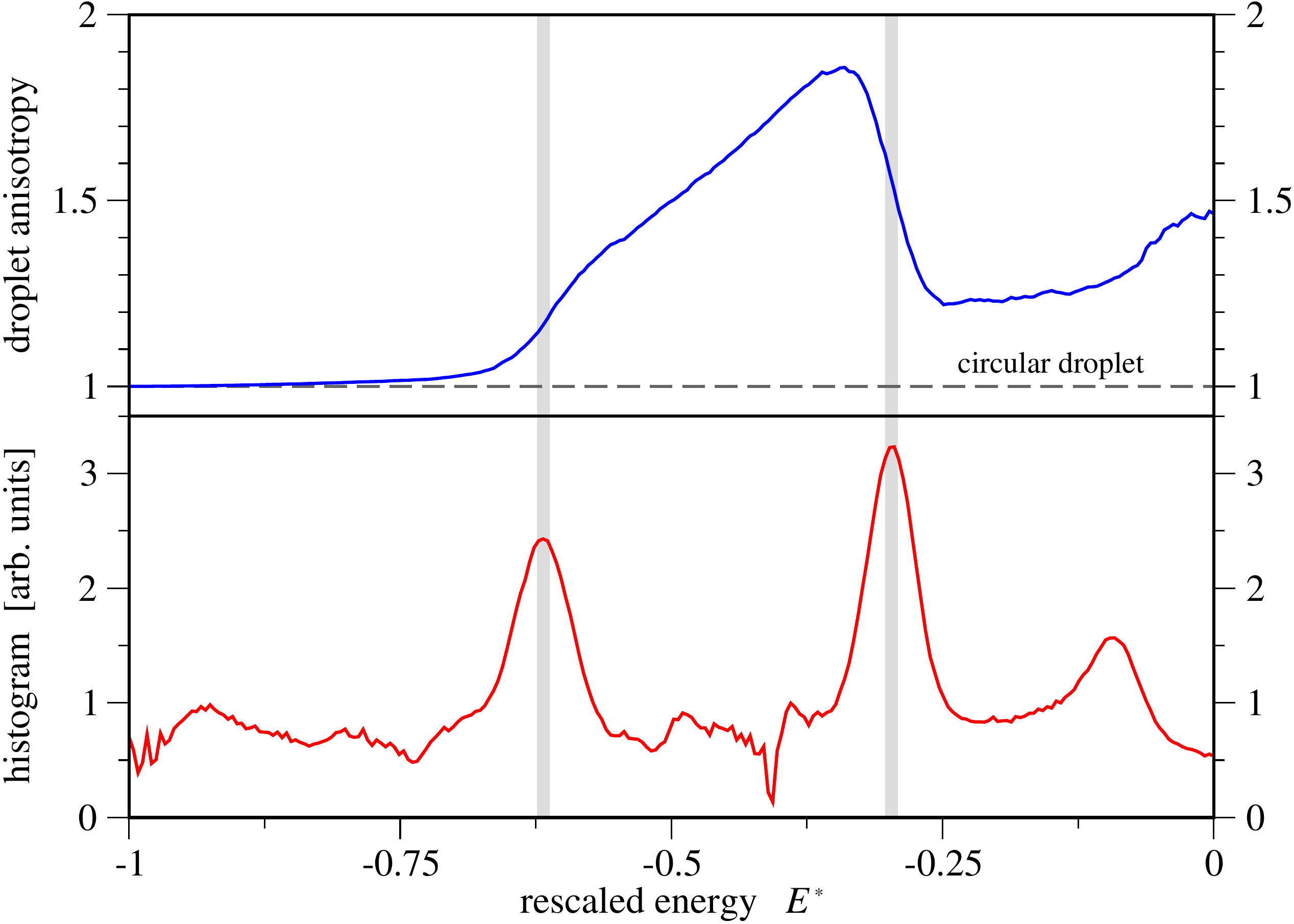}
	\caption{(color online)
	                Anisotropy of the largest droplet of ordered phase as defined in
	                Eq.~\eqref{eqn:anisotropy}
	                in the phase coexistence region of a 250-state Potts model ($L=18$).
                         A spherical droplet corresponds to anisotropy 1 as indicated by the dashed line.
                         At the droplet-strip transitions indicated by the vertical bars the droplet anisotropy
	                quickly changes.
	                The deviation from unity above the upper droplet-strip transition indicates an
	                anisotropy due to finite-size effects.}
	\label{fig:anisotropy}
\end{figure}

Finally, we return to the droplet anisotropy and estimate corrections to the circular shape induced by the finite size of our system. To this end we monitor the anisotropy of the droplet by measuring the ratio of the dimensions $L_1, L_2$ of the minimal bounding box for the largest (ordered) droplet in the system
\begin{equation}
   a = \frac{\max(L_1, L_2)}{\min(L_1, L_2)} \,.
   \label{eqn:anisotropy}
\end{equation}
In this notation a spherical droplet corresponds to $a = 1$.

In Fig.~\ref{fig:anisotropy} we plot this anisotropy for the largest ordered droplet measured for energies in the phase coexistence region.
At the droplet-strip transitions -- indicated by the vertical bars in the figure --
the droplet anisotropy undergoes rapid changes as the droplet percolates and deforms into a strip.
For energies above the upper droplet-strip transition we find that the droplet anisotropy deviates from unity which indicates that the ordered droplet in an otherwise disordered phase exhibits an non-spherical rather than a spherical shape. Since the analytical calculation for a droplet embedded in an infinite system predicts a roughly spherical shape~\cite{fujimoto1997}, the observed anisotropy must be rooted in the finite size of our system. Interestingly, we also seem to observe a signature indicating the droplet formation with the droplet anisotropy suddenly increasing from $a \approx 1.25$ to $a \approx 1.5$ in the respective energy region.

\subsection{Simulations on the cube surface}

\begin{figure}[t]
	\includegraphics[width=\columnwidth]{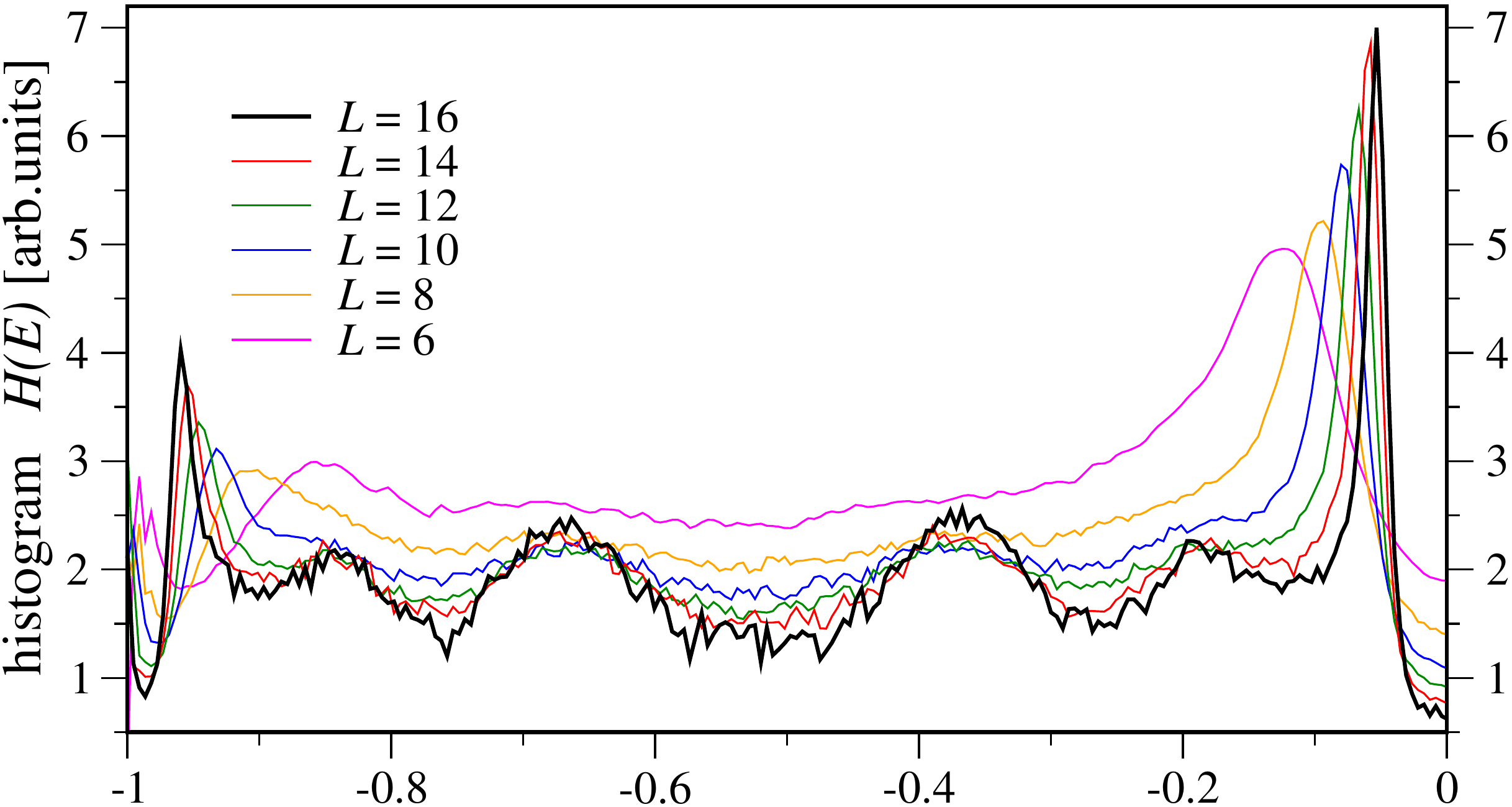} \\
	\includegraphics[width=\columnwidth]{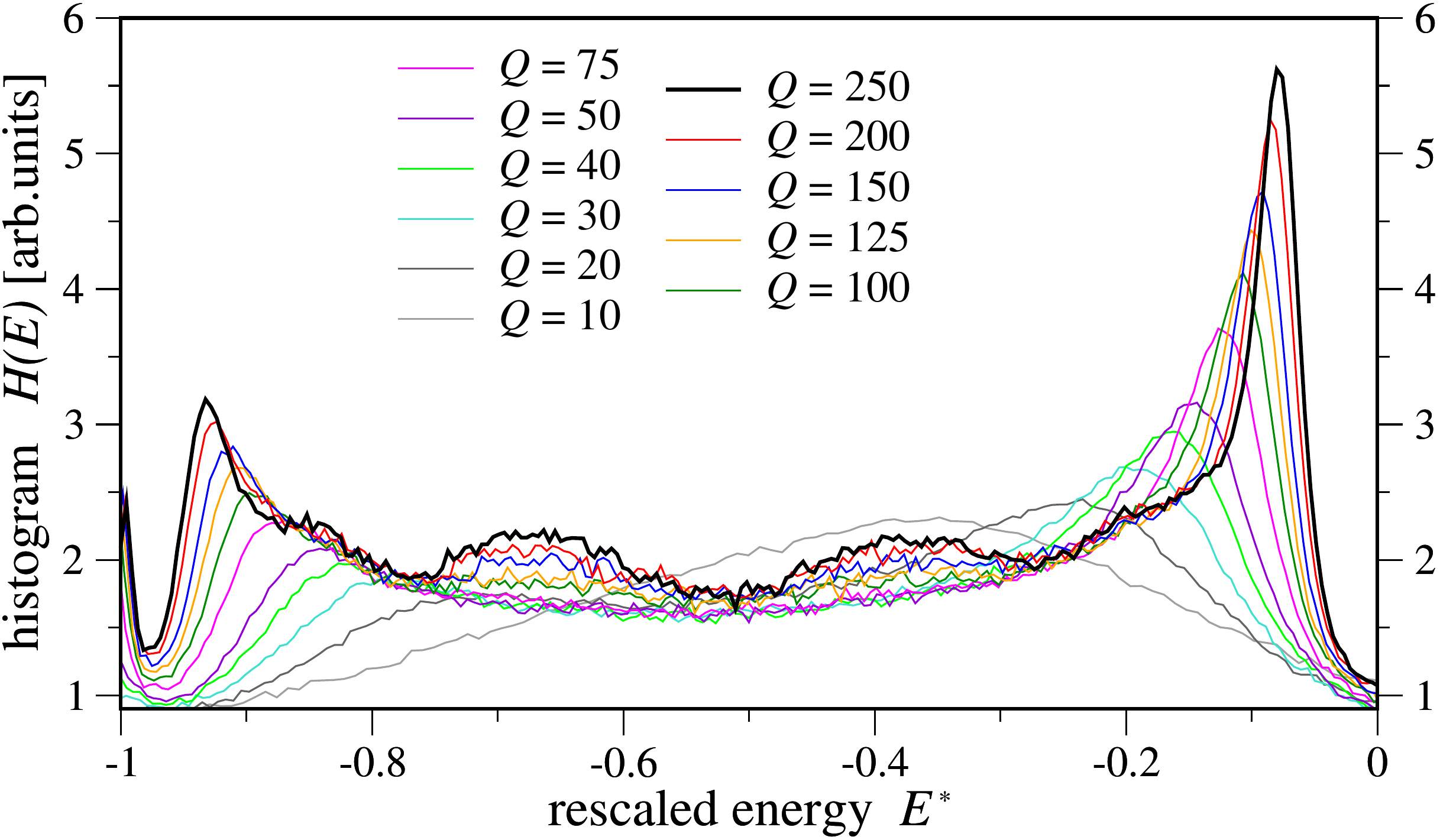}
	\caption{(color online)
		Optimized histogram for the Potts model on the cube surface
		for (a) a fixed number of Potts states $Q = 250$ and various system sizes $L$,
		(b) a fixed system size $L=10$ and
		different choices of the number of Potts states $Q$.
		In contrast to the simulations on the torus (cf. Fig.~\ref{fig:histL}),
		the dominant peaks are related to droplet nucleation/annihilation
		transitions, while the emerging, smaller peaks reveal transitions
		between states with droplets occupying an increasing number of corners
		of the cube.
	         The plotted energy range corresponds to the coexistence region defined in
		Eq.~\eqref{eqn:rescaling}.
	}
	\label{fig:histL_cube}
\end{figure}

Since the droplet-strip transition causes the main bottleneck for simulations in a toroidal geometry,
one could ask whether simulations on other surface topologies, in particular on simply-connected surfaces,
do not suffer from entropic barriers originating from such shape transitions.
In order to explore this idea we have simulated the Potts model in a ``sphere topology" by
considering the surface of a cube. We assemble such a cube surface with $L$ sites on each edge and a total
number of sites $N = 6 L^2-12 L+8$ such that the corners have coordination number $z=3$,
while all other sites have coordination number $z=4$.

\begin{figure}[t]
	\includegraphics[width=\columnwidth]{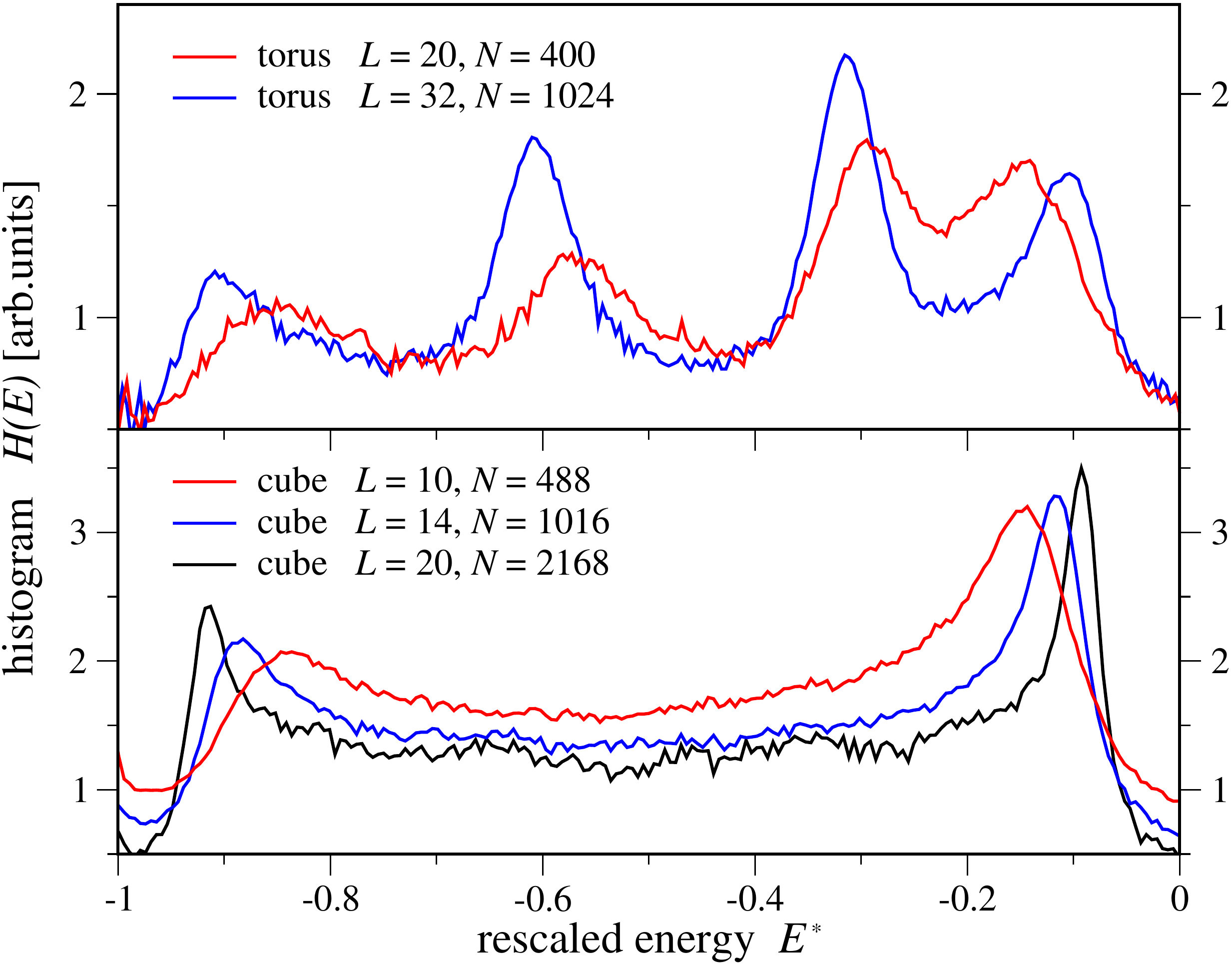}
	\caption{(color online)
		Comparison between optimized histograms for the cube surface and the torus at $Q=50$
		for various system sizes $L$.
	         The plotted energy range corresponds to the coexistence region defined in
		Eq.~\eqref{eqn:rescaling}.
	}
	\label{fig:torus_cube_comparison}
\end{figure}

Fig.~\ref{fig:histL_cube} shows results for the optimized histograms for the Potts model on such a cube surface. In panel a), histograms for fixed $Q=250$ and $L$ in the range $L = 6 \ldots 16$ are shown, while in panel b), the system size is fixed at $L = 10$ and simulations are shown for various $Q = 10 \ldots 250$. As opposed to the simulations on the torus, the peaks corresponding to droplet nucleation/evaporation are now dominating the histogram for these parameters.

However, for the largest systems with $Q = 250$ and $L \geq 14$ ($N \geq 1016$ spins) spins, four smaller peaks emerge. Examining snapshots of the system in the associated energy ranges, we find that droplets nucleate around corners
of the cube and the transitions mark a change in the number of occupied corners. A similar observation has been made previously for multicanonical simulations of the Ising model in an external field~\cite{neuhaus2003}.

There are multiple reasons why droplets nucleate at the corners of the cube. Naively, one might think that this is
solely due to the lower coordination number of the corner sites, which provides a small energetic advantage
(of order one in the system size) to place a droplet on a corner.
Closer inspection of the surface free energy of a droplet enclosing a corner, however, reveals that similar to the
droplet-strip transition there are entropic barriers which scale with the system size $L$.
To see this consider a droplet of area $A$.  If $A$ is small enough (relative to $L^2$), this droplet may sit on one of the faces of the cube so that it encloses no corners of the cube. Then the surface it sits on is flat, so the surface free energy is minimized by a circular droplet with radius $R$ so that $A=\pi R^2$. Alternatively, the drop may enclose one corner.  Putting the corner at the center of the droplet, the droplet can be a quarter-circle on each of the 3 adjacent faces.  These quarter-circles each have radius $\tilde{R}$ with $A=(3\pi \tilde{R}^2)/4$.  The net result is that the perimeter of the droplet (which sets the surface free energy) is smaller by a factor of $\sqrt{3}/2$ when the droplet is centered on a corner as compared to when it does not contain a corner.  The difference is of order $\sqrt{A}$, so for large enough cubes will dominate over the order one effect mentioned above.

While the simulation bottlenecks / entropic barriers associated with the corner transitions are significantly
suppressed in comparison with the droplet-strip transitions of the toroidal geometry, which is illustrated
Fig.~\ref{fig:torus_cube_comparison}, we have not succeeded in suppressing {\em all} entropic barriers of
order $L$ by going from the torus to the cube surface. As a consequence, we expect the same asymptotic
performance of the optimized ensemble simulations for both geometries.

%%%%%%%%%%%%%%%%%%%%%%%%%%%%%%%%%%%%%%%%%%%%%%%%%%%
\section{Sampling efficiency and round-trip times}
\label{sec:performance}

\begin{figure}[t]
	\includegraphics[width=\columnwidth]{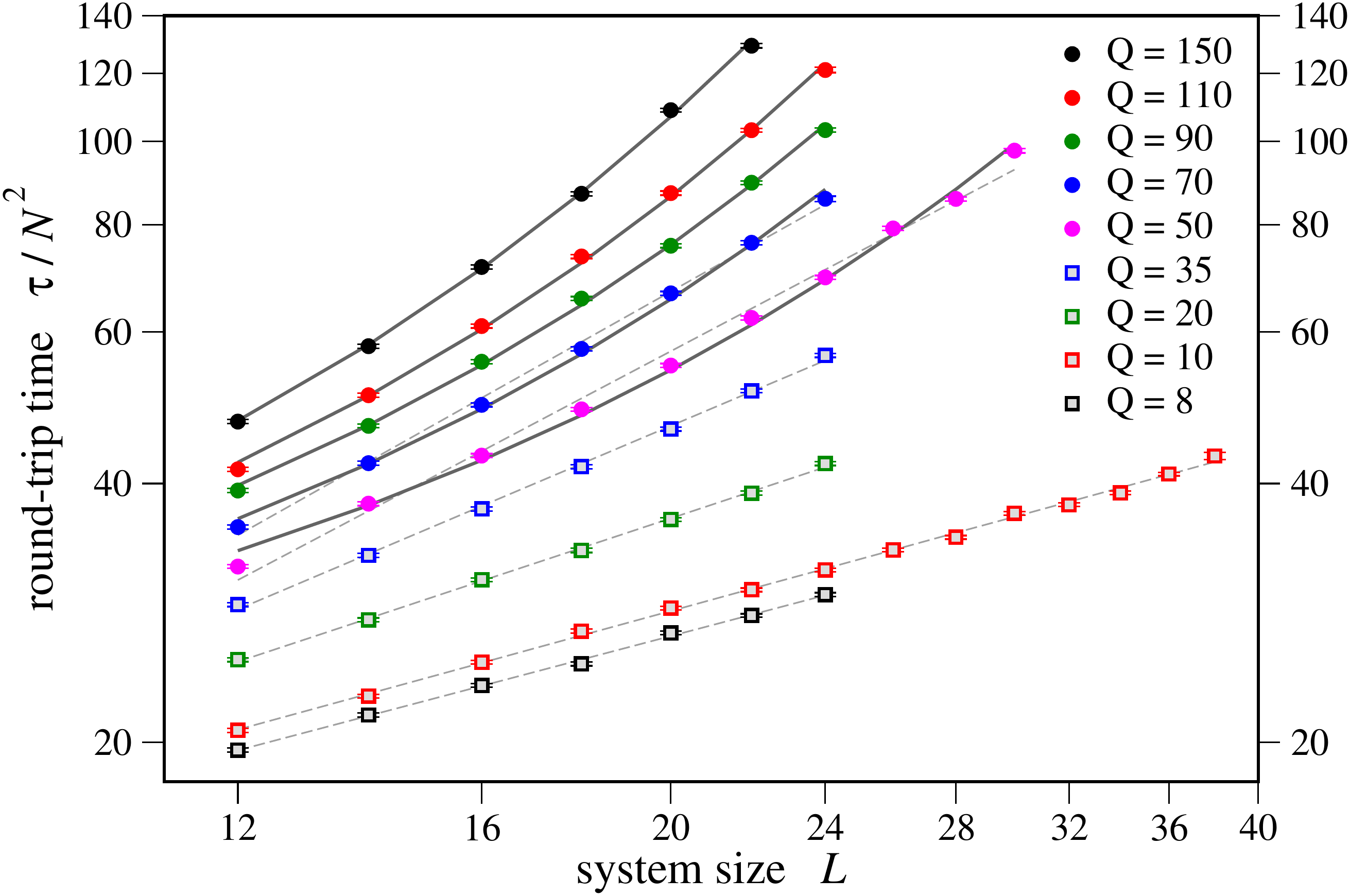}
	\caption{(color online)
	                Scaling of the round-trip time between the lowest and highest energy states with
	                system size $L$ for the optimized ensemble simulations using heat bath dynamics and the toroidal geometry.
	                While the scaling can be fitted to polynomial behavior for small $Q \lesssim 35$,
	                the scaling crosses over to exponential behavior for larger $Q \gtrsim 50$.}
	\label{fig:scalingL}
\end{figure}

\begin{figure}[t]
	\includegraphics[width= \columnwidth]{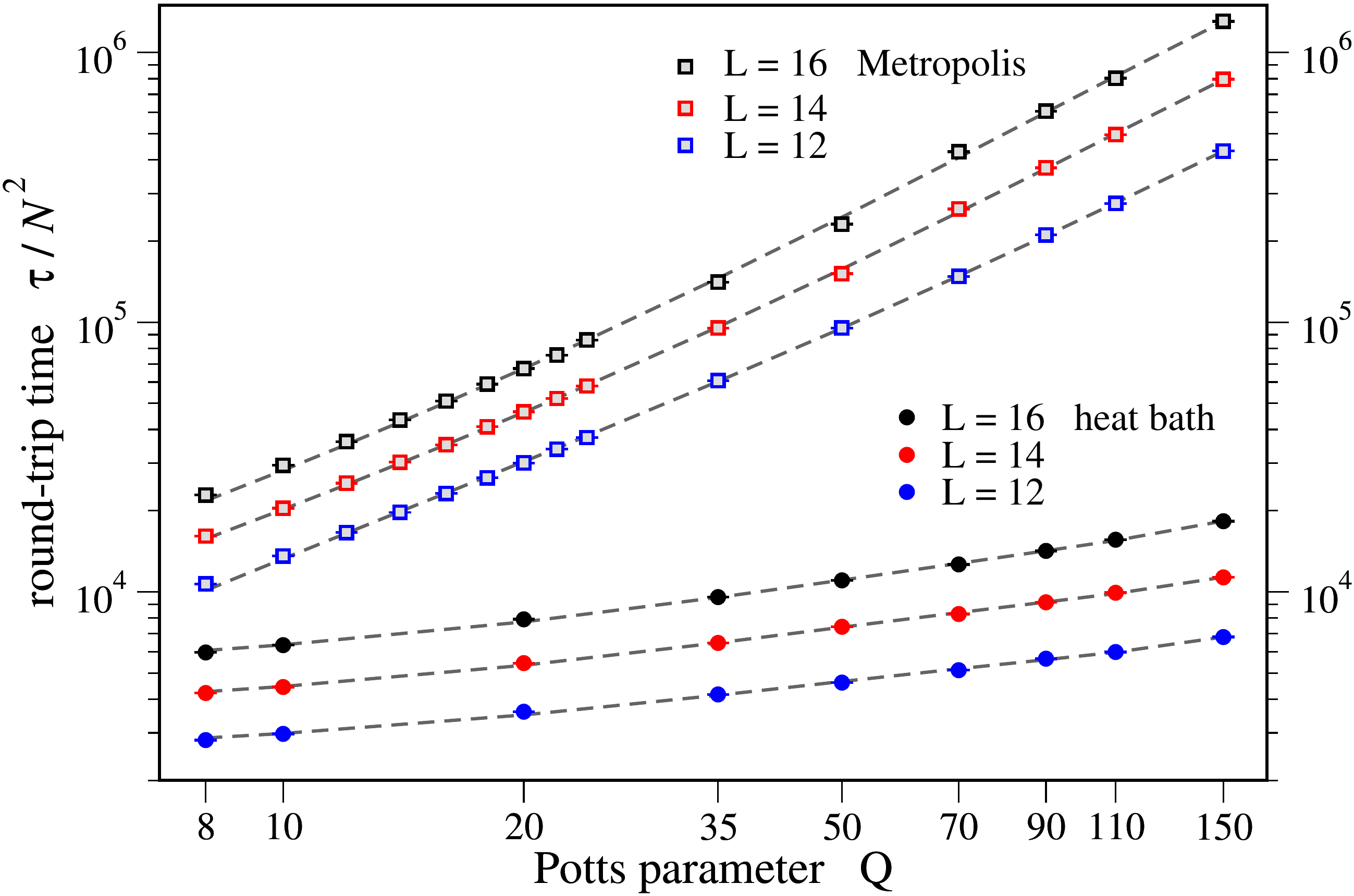}
	\caption{(color online)
	                Scaling of the round-trip time between the lowest and highest energy states with
		       the number of Potts states $Q$ for  the optimized ensemble simulations using both
		       Metropolis and heat bath dynamics in the toroidal geometry.
		      The dashed lines represent cubic regressions to the data.}
	\label{fig:scalingQ}
\end{figure}

We finally address the numerical efficiency of sampling the optimized statistical ensemble
for the $Q$-state Potts model and, more generically, for a strong first-order phase transition.
In order to quantify this performance we follow earlier work~\cite{dayal2004,trebst2004,alder2004}
and measure the characteristic time scale of the random walk in energy space, e.g. the round-trip time
to traverse the full energy range $[E_-, E_+]$, and its scaling with system size $L$ and number of Potts states $Q$.
Our results are summarized in Figs.~\ref{fig:scalingL} and \ref{fig:scalingQ}.

For systems undergoing continuous transitions it was shown~\cite{dayal2004} that the flat-histogram ensemble sampled in the Wang-Landau method %does not produce this perfect scaling, but 
generally
suffers from a power-law slow down
\begin{equation}
  \tau_{\rm flat-histogram} \propto N^2 \cdot N^z \,,
\end{equation}
which is  reminiscent of the well-known critical slowing down in the canonical ensemble.
The additional exponent $z$ depends on the system at hand, and was measured to be $z \approx 0.4$ for the Ising model and $z \approx 0.9$ for the fully frustrated Ising model~\cite{dayal2004}.
In contrast, the optimized ensemble method does not suffer from such a `critical slowing down'
at continuous transitions and produces round-trip times that scale almost perfectly with system size
\begin{equation}
  \tau_{\rm optimized-ensemble} \propto N^2 \cdot (\ln N)^2 \,,
  \label{eq:optimized-scaling}
\end{equation}
up to a logarithmic correction~\cite{trebst2004}.
The improved scaling of the optimized ensemble can thus considerably speed up the sampling efficiency of a broad-histogram simulation, with improvements of two orders of magnitude reported already for intermediate system sizes~\cite{trebst2004}.

Exploring the efficiency of simulations at a first-order phase transition, previous work has found an exponential divergence of the round-trip time with $L$ in the Ising model in an external magnetic field~\cite{neuhaus2003}, which is referred to as \emph{supercritical slowing down}. This divergence occurs in the presence of shape transitions such as the droplet-strip transition discussed above and the quantitative behavior of the scaling can indeed be related to the surface tension of droplets.
In a recent study Neuhaus {\em {\it et al.}.}~\cite{neuhaus2007} demonstrated that the multiple Gaussian modified ensemble (MGME) method suffers only from a residual supercritical slowing down when applied to the Potts model
(with $Q = 20$ and $Q = 256$ states in their simulations).

For the optimized ensembles the scaling of the round-trip times is shown in  Figs.~\ref{fig:scalingL} and \ref{fig:scalingQ} as a function of both system size $L$ and number of Potts states $Q$, respectively.
Similar to previous results for continuous transitions, we find a dramatic improvement of the optimized ensemble when compared to the flat-histogram ensemble. However, even for the optimized ensemble we do not recover the almost perfect scaling \eqref{eq:optimized-scaling} observed for continuous transitions when the severity of the first-order transition proliferates, e.g. by increasing the system size $L$ or the number of Potts states $Q$.
In general, we find that the scaling is independent of the transition dynamics, e.g. whether we choose Metropolis or the heat-bath transition rates.

We first look at the scaling with system size $L$ when fixing the number of Potts states $Q$ in the range $8 \leq Q \leq 150$. As shown in Fig.~\ref{fig:scalingL} we find that, for the range of $L$ studied, the round-trip times appear to scale polynomially $\tau \sim N^{2+z}$ in system size $L$ for $Q \lesssim 35$. We estimate the effective exponents to be  $z \approx 0.3$ for $Q = 8$, $z \approx 0.31$ for $Q=10$, $z \approx 0.38$ for $Q=20$, and $z \approx 0.48$ for $Q=35$.
We note that in this regime the first-order transition is still relatively weak and we do not (yet) observe the characteristic multi-peak structure in the optimized histogram. However, when further increasing number of Potts states $Q$ we find that the scaling crosses over to the expected exponential behavior $\tau \sim \exp(\alpha L)$ for $Q \gtrsim 50$ due to the entropy barriers at the droplet-strip transitions, and this becomes apparent already on rather small system sizes. This, of course, bears witness of the strong first-order nature of the phase transition in this regime, which also becomes evident in a noticeable multi-peak structure of the optimized histogram even for considerably small system sizes $L \geq 14$. Our results however do not exclude a crossover to supercritical scaling also for weaker transitions at some larger system size. Indeed, results in Ref.~\onlinecite{mayor2007} indicate that also for $Q = 10$, an extended stripe phase emerges for system sizes $L \gtrsim 512$, which may well lead to the same effect we observe for large $Q$ at much smaller $L$.

%%%%%%%%%%%%%%%%%%%%%%%%%%%%%%%%%%%%%%%%%%%%%%%%%%%
% Conclusions ---
%%%%%%%%%%%%%%%%%%%%%%%%%%%%%%%%%%%%%%%%%%%%%%%%%%%

\section{Conclusions}

We presented an application of the optimized ensemble method to improve equilibration
of broad-histogram simulations of the first-order transition in the large-$Q$ Potts model.
The optimized histogram develops a characteristic multipeak structure which indicates
that the system releases latent heat in a sequence of phase transitions.
The intermediate metastable states exhibit droplets of the coexisting phases in varying shapes.
For a toroidal system geometry the dominant bottleneck in the simulation is the entropic barrier
associated with a droplet-strip transition.

We find that the ensemble optimization is capable to only
partially overcome this bottleneck by shuffling statistical weight towards the entropic barriers.
It still exhibits the same asymptotic supercritical slowing down for strongly first-order transitions previously
reported for flat-histogram simulations~\cite{neuhaus2003}.
It thus remains an open question whether simulation schemes for strongly first-order phase transitions 
can be further improved to overcome this asymptotic slowing down. 
Possibly, statistical ensembles defined in multiple system variables 
could further improve extended ensemble simulations, or one might turn to modified update technique, 
which, for instance, attempt to specifically update the boundaries of a droplet.

%%%%%%%%%%%%%%%%%%%%%%%%%%%%%%%%%%%%%%%%%%%%%%%%%%%
\section*{Acknowledgments}
Our numerical calculations were performed on the Hreidar and Brutus clusters at ETH Zurich.
S.~T. thanks the MPI-PKS Dresden for hospitality.

\bibliography{bibliography}{}
\bibliographystyle{apsrev}

\end{document}